\documentclass[journal,comsoc]{IEEEtran}

\usepackage[T1]{fontenc}

\ifCLASSINFOpdf
   \usepackage[pdftex]{graphicx}

  \DeclareGraphicsExtensions{.pdf,.jpeg,.png}
\else

\fi
\usepackage{amsmath}

\usepackage[cmintegrals]{newtxmath}
\usepackage[caption=false]{subfig}

\usepackage{graphicx}
\usepackage{latexsym}
\usepackage{amsmath,epsfig}
\usepackage{epstopdf}
\usepackage{caption}
\usepackage{multirow}
\usepackage{booktabs,siunitx}
\usepackage[utf8]{inputenc}
\usepackage[T1]{fontenc}
\usepackage{lscape}
\usepackage{url}
\usepackage{graphicx}
\usepackage{epsfig}
\usepackage{graphicx}
\usepackage{latexsym}
\usepackage{amsmath}
\usepackage{amssymb}
\usepackage{mhsetup}
\usepackage{mathtools}
\usepackage{mathrsfs}
\usepackage{float}
\usepackage{xfrac}
\usepackage{epsfig}
\usepackage[english]{babel}
\usepackage{subfig}
\usepackage{algorithm}
\usepackage[noend]{algpseudocode}

\begin{document}

\title{Accurate, Energy-Efficient, Decentralized, Single-Hop, Asynchronous Time Synchronization Protocols for Wireless Sensor Networks}

\author{Ramadan~Abdul-Rashid,~\IEEEmembership{}
	Ali~Al-Shaikhi,~\IEEEmembership{Member,~IEEE,}
	and~Ahmad~Masoud,~\IEEEmembership{Member,~IEEE}
\thanks{R. A. Rashid is a  graduate student in the Electrical Engineering Department of King Fahd University of Petroleum and Minerals, Dhahran, 31261, Saudi Arabia e-mail: ram.rashid.rr@gmail.com}
\thanks{Dr. A. Al-Shaikhi is an Assistant Professor with the Electrical Engineering Department of King Fahd University of Petroleum and Minerals, Dhahran, 31261, Saudi Arabia e-mail: shaikhi@kfupm.edu.sa}
\thanks{Dr. A. Masoud is an Associate Professor with the Electrical Engineering Department of King Fahd University of Petroleum and Minerals, Dhahran, 31261, Saudi Arabia e-mail: masoud@kfupm.edu.sa}%

\thanks{Manuscript received on; revised }}

\maketitle

\begin{abstract}
This paper concerns with the synchronization of infrastructure impoverished sensor networks under harsh conditions. It suggests three novel asynchronous, decentralized, energy-efficient time synchronization protocols. The protocols require only single hop, sparse communication with unlabeled neighboring nodes of the network to determine accurately the time of the gateway node. The time of a node is considered as a dynamical variable of a discrete system whose evolution is asynchronously activated/inhibited by another dynamical switching system. The protocols are termed:  Timed Sequential Asynchronous Update (TSAU), Unidirectional Asynchronous Flooding (UAF) and the Bidirectional Asynchronous Flooding (UAF). Along with intensive simulation, the protocols are implemented and tested on the MicaZ sensor node platform. A comprehensive evaluation of the energy consumption, memory requirements, convergence time, local and global synchronization errors of the proposed protocols are carried-out against Flooding Time Synchronization Protocol (FTSP) and Flooding Proportional Integral Time Synchronization Protocol (FloodPISync). All the analysis show that these protocols outperform the well known protocols. Also, being asynchronous, they are more realistic relative to the synchronous ones.
\end{abstract}


\IEEEpeerreviewmaketitle

\section{Introduction}
Wireless Sensor Networks (WSNs) are distributed systems used for several sensing and instrumentation applications. They are normally deployed under harsh conditions such as in experimental environments with extremely varying  temperatures, heavy storms, vibrations, concentrated chemicals, electrical shocks, and mechanical pressures \cite{_wiley:_2015,_wireless_2015}.

The nodes of a WSN are required to communicate their time-stamped, sensed data to the gateway node. This, in turn, requires the WSN nodes to be closely synchronized to the time of that node. Several factors have to be addressed in order to achieve accurate and robust time synchronization for a WSN \cite{jeremy_elson_wireless_2003}. Some of these factors are: scarcity of energy of the deployed nodes, scalability and the need for large scale WSNs, decentralized topologies, unpredictable and intermittent connectivity between network nodes. The constraints of limited resources and operation in harsh environments requires time synchronization algorithms operating on a WSN to be simple, efficient, robust and have high rate of convergence so as to keep network nodes synchronized on-demand or at all times. A myriad of studies has gone into developing different methods of time synchronization for WSNs with the goal of optimizing networks parameters like energy utilization, precision, lifetime, scope, scalability, and cost.

The development of time synchronization algorithms began, for the most part, with the development of centralized protocols \cite{Jiming_2010,kadowaki_event-based_2015}. These schemes achieve global network time-synchronization by synchronizing all network nodes to a root/leader/reference node in a hierarchical tree fashion or using network clusters. Popular among these protocols are the Reference Broadcast Synchronization (RBS) \cite{elson_fine-grained_2002}, Timing-synch Protocol for Sensor Networks (TPSN) \cite{ganeriwal_timing-sync_2003}, Lightweight Time Synchronization (LTS) \cite{dai_tsync:_2004}, Hierarchical Time Synchronization (HRTS) \cite{van_greunen_lightweight_2003}, Flooding Time Synchronization Protocol (FTSP) \cite{maroti_flooding_2004}, and Flooding Proportional Integral Time Synchronization Protocol (FloodPISync) \cite{yildirim2018adaptive}. Centralized protocols are not suited for use with WSNs where below average hardware, node failure, changing network topology, scarce resources and intermittent connectivity are major issues.

Distributed synchronization schemes have better ability to address the difficulties faced by centralized methods \cite{he_time_2014}.  Most of the distributed synchronization algorithms are based on consensus. Consensus based methods do not require a root (leader) node or a spanning tree for time synchronization. They assume that the time of each node of the WSN is parameterized by offset and skew. The algorithm in each node attempts to reach an agreement with the others regarding the values these two parameters should attain in order to synchronize the network. Key examples of the consensus protocols are:  Average Proportional Integral Time Synchronization Protocol (AvgPISync) \cite{yildirim2018adaptive}, Gradient Time Synchronization Protocol \cite{sommer_gradient_2009}, and Average Time Synch (ATS) Protocol \cite{schenato_average_2011}. Other distributed synchronization protocols are: Time Synchronization Protocol using the Maximum And Average Values (TSMA) \cite{dengchang_time_2013}, Time Diffusion Protocol (TDP)\cite{su_time-diffusion_2005}, External Gradient Time Synchronization Protocol (EGSync) \cite{yildirim_external_2014}, Reach-Back Firefly Algorithm (RFA) \cite{werner-allen_firefly-inspired_2005}, Gradient Time Synchronization Protocol(GTSP) \cite{sommer_gradient_2009}, Energy Efficient Gradient Time Synchronization Protocol(EGTSP), Adaptive Value Tracking Synchronization (AVTS) Protocol, Average Time-Sync Protocol\cite{schenato_average_2011}, and Weighted Maximum Time Synchronization Protocol(WMTS) \cite{he_time_2014}.  Although distributed protocols address to a certain extent the shortcomings of centralized protocols, they still suffer from issues that prevent them from reliably operating in harsh environments. For example, in ATS and GTSP, communication delays are not taken into consideration in the protocol design. Other distributed protocols such as AVTS and AvgPISync consider communication delays. However, communication overhead required for synchronization tends to be high since the nodes must constantly keep transmitting and receiving synchronization messages within a short time interval across multiple hops. This makes the process susceptible to node and communication link failures. In addition, energy consumption is high.  Other distributed methods such as WMTS and EGTSP do not consider convergence time in protocol design and evaluation. Moreover, the synchronization accuracy at convergence is relatively low. This causes nodes’ times to drift quickly prompting frequent resynchronization. The above issues makes it difficult for a large-scale WSNs built from inexpensive, scattered nodes and operating in harsh environments to function reliably. In general, there is still a need for reliable and accurate distributed synchronization procedures that suit networks comprised of inexpensive nodes with limited resources operating in harsh environments.

This paper suggests an efficient, decentralized synchronization protocol that suits an infrastructure-impoverished WSN operating in a harsh environment. Communication among the nodes of the network is carried-out in an event-based asynchronous manner. The suggested approach departs from existing decentralized approaches by assuming that local time is a free-evolving dynamical variable that is not parameterized by drift and skew. Each node, independently, when triggering event is true, evolves the value of that variable using the time variables of its single hop neighbors. The protocol can determine from only its own local time variables which communication instant is closest to the time of the gateway node. The value of the variable at that instant is used to initialize the physical clock of the node. It is observed that local determination of the instant at which local time of a node is closest to the gateway time can be done at high accuracy with small communication overhead. The decentralized synchronizing paradigm is developed and three realizations of this paradigm are provided. They are termed: Timed Sequential Asynchronous Update (TSAU), Unidirectional Asynchronous Flooding (UAF) and the Bidirectional Asynchronous Flooding (UAF). The realizations are thoroughly simulated and experimented with the MicaZ WSN platform. Experimental comparisons are conducted between the suggested and existing synchronization protocols. The suggested asynchronous paradigm builds on the work in \cite{al2017efficient} that assumes network-wide synchronous communication. It is worth mentioning that the suggested asynchronous paradigm uses the gateway node just as the reference time to which all nodes in the network need to synchronize to. This asynchronous paradigm for achieving consensus is similar to asynchronous algorithms like the pairwise and neighborhood gossip \cite{denantes_which_2008} and the geographic gossip \cite{kar_distributed_2009} but has a unique formulation as compared to these traditional asynchronous algorithms.

The rest of the paper is organized as follows. Section~\ref{The Synchronization Procedure} provides a quick background review of the synchronous protocol on which the suggested asynchronous protocol is based. Section~\ref{The Proposed Synchronizer - A General Setup} provides the general setup of the suggested synchronizer. Section~\ref{Realizations of the Proposed Asynchronous Paradigm} presents the three realizations of the synchronization method. Sections~\ref{Simulation Evaluation}~and~\ref{Real-Time Experimental Evaluation} contain extensive simulations, experiments, and evaluations of the suggested procedures. The final findings of the paper is summarized in section~\ref{Conclusion}.

\section{The Synchronization Procedure}
\label{The Synchronization Procedure}
To meet the challenges a harsh environment imposes, the suggested synchronizer exploits the synergy between the hardware and the software parts of the sensor network. The nodes are programmed so that the overall network forms a discrete dynamical system whose states represent the time estimate each node has for the time of the gateway node. All the nodes use identical programs for updating their local time estimates (i.e. protocols). These programs require input data only from the immediate neighbors of a node. They are executed in an asynchronous manner that depend only on their local states and the local data they acquire. As a whole, the network may be represented as a discrete, switched dynamical system \cite{lin2009stability},

\begin{equation}\label{eq1}
T_{k} = A_{\sigma}T_{k-1} + B_{\sigma}t_{gk}
\end{equation}

where, $T_k$ is a vector whose elements are the evolving time series which each node uses to determine the gateway time. $A_{\sigma}$  is an $(N-1)\times (N-1)$ matrix representing a switched connectivity matrix which is controlled by a switching rule $\sigma$ of a WSN \cite{colaneri2015analysis,lin2009stability} with $N-1$ nodes and 1 gateway node and $B_{\sigma}$ is an $(N-1)\time1$ vector whose structure changes based on a switching rule $\sigma$ and $t_{gk}$ represents the evolving gateway time.

Each node $i$ updates its time $t_{i}$ at each communication instant $k$ based on a sub-dynamical system:
\begin{equation}\label{eq2}
t_{ik} = a\sigma_{i} t_{i(k-1)} + b\sigma_{i} t_{gk}
\end{equation}

\begin{figure}[h]
	\centering
	\includegraphics[width=0.45\textwidth]{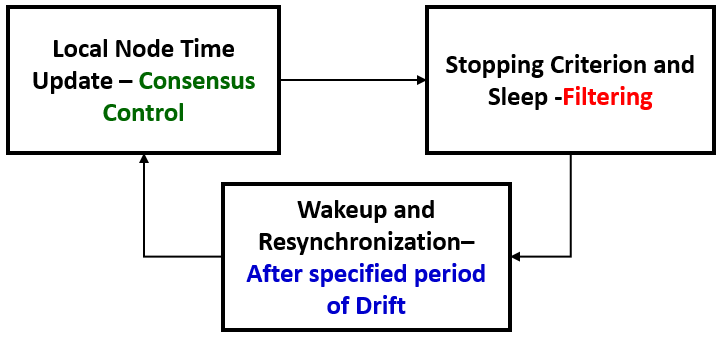}
	\caption[Block Diagram of Suggested Synchronization Procedure]{Block Diagram of Suggested Synchronization Procedure}
	\label{fig:pro_block}
\end{figure}

Each discrete instant in time, $k$, represents a communication attempt among the nodes and cause an energy drain that cannot be neglected in the operation of the network. Incurring an excessive communication overhead by having to wait until the system reaches steady state is not practical. The suggested protocol solves this problem by proposing a method to stop node update while the system is in the transient phase. As will be demonstrated, terminating operation in the transient phase does not only save communication energy, but also significantly increases the accuracy of the synchronization procedure. Moreover, all nodes terminate updating their times, independently, at approximately the same time instant. This makes it possible to utilize the knowledge about the quality of the hardware used and how fast the clock of a node drifts to schedule sleep and wake-up periods of the network. The work-flow of the synchronizer is shown in Figure~\ref{fig:pro_block}. Each component of the synchronizer is discussed next.

\subsection{Node Local Time Update}
In~\cite{al2017efficient}, a synchronous average consensus approach is adopted for local time update at each network node. This approach assumes a simultaneous update for all network nodes at each communication round as described below.

\begin{figure}[htbp!]
	\centering
	\includegraphics[width=0.5\textwidth]{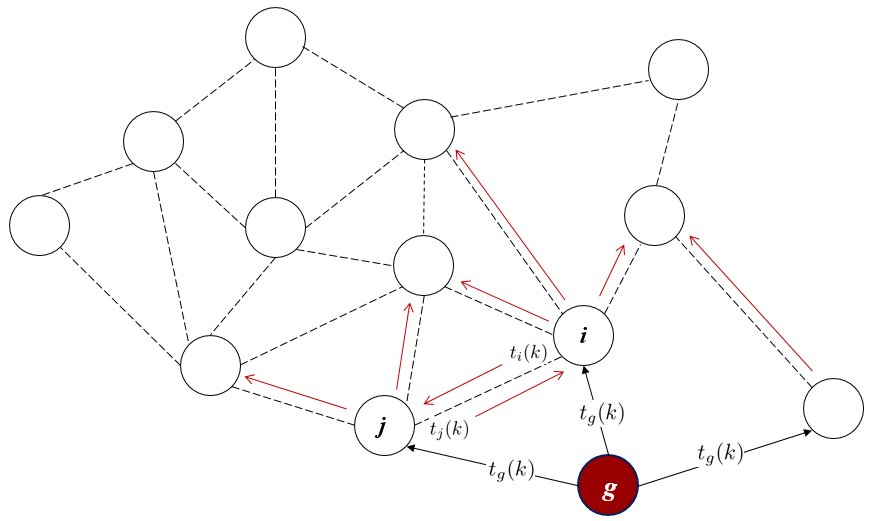}
	\caption[Network of N Nodes and L Links]{Network of N Nodes and L Links}
	\label{fig:net}
\end{figure}

Consider a network represented by a graph $\mathcal{G} = (V,E)$, where $V = \{1,\hdots ,N\}$ is the vertex set representing nodes in a wireless sensor network, with 1 gateway node, $g$ and $N-1$ ordinary nodes and the network connectivity between these nodes as an edge set, $E$. Assuming a node $i\in V$ can exchange its logical clock information with node $j$ in a single-hop, i.e. $(i,j)\in E$ and $j\in \mathcal{N}_{i}$, where $\mathcal{N}_{i}$ describes the neighborhood set of node $i$ as shown in Figure \ref{fig:net}. If there exists a spanning tree covering each node of the WSN with a root node as the gateway then a node $i$ can use the average of the times of its immediate single hope neighbors to drive it’s time estimate $t_i(k)$ arbitrarily close to the time of the gateway node $t_g(k)$ \cite{al2017efficient}.

At the $k_{th}$ communication time instant, when a node $i$ receives the times from the set of its single hop neighbors $(\mathcal{N}_i)$, it computes its time estimate $t_i(k)$ by averaging the neighbors time $(t_j(k), j=1, \hdots ,|\mathcal{N}_i|)$ using the formula:

\begin{equation}\label{eq3}
t_i(k) = \frac{1}{|\mathcal{N}_{i}|} \sum_{j \in \mathcal{N}_i}^{}t_j(k-1)
\end{equation}
where $|\mathcal{N}_{i}|$ denotes the neighborhood cardinality of $|\mathcal{N}_{i}|$.
This synchronization procedure assumes that the gateway node ticks as a perfect clock given by $ t_g(k) = \Delta k$, where $\Delta$ is the ticking rate of the gateway clock. If $i$ is also connected to the gateway node, then the update is expressed in terms of both neighbors' time estimate and gateway clock as:
\begin{equation}\label{eq4}
t_i(k) = \frac{1}{|\mathcal{N}_{i}|} \sum_{j \in \mathcal{N}_i}^{} \left[ t_j(k-1) + \Delta k \right]
\end{equation}

\subsection{Dip Phenomenon and Stopping Criterion}
\begin{figure}[htbp!]
	\centering
	\includegraphics[width=0.5\textwidth]{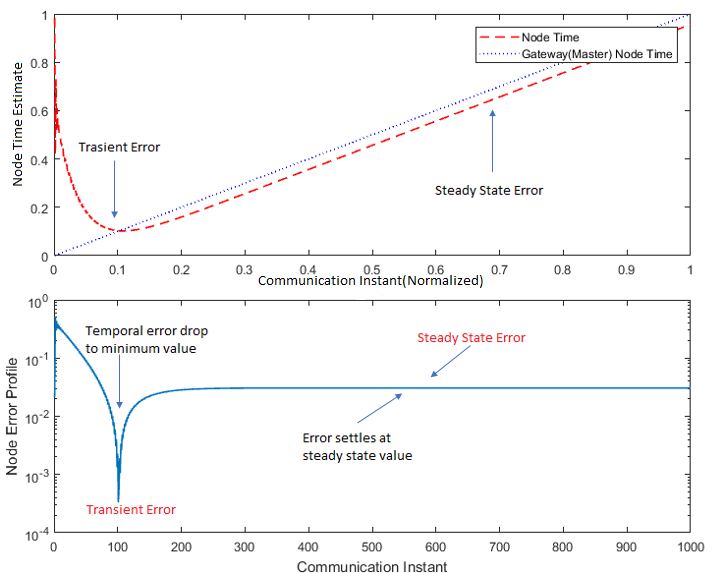}
	\caption[Transient (Dip) and Steady State Behavior of Node Clocks with respect to Gateway Node Clock]{Transient (Dip) and Steady State Behavior of Node Clocks with respect to Gateway Node Clock}
	\label{fig:errorP}
\end{figure}

It is shown in~\cite{al2017efficient} that the procedure represented by (\ref{eq1}) can drive all network nodes' times to the clock of the gateway node with a very small error margin if the rate of the gateway clock is significantly reduced. This solution is not practical since an increase in the gateway clock ticking rate would mean a higher number of communication cycles leading to high energy consumption~\cite{al2017efficient}. It is observed, however, from the evolution of node local clock value with respect to the gateway clock that there exists a certain time estimate, $t_i(k)$, where the node's local time curve intersects with that of the gateway clock curve as illustrated by Figure~\ref{fig:errorP}. This phenomenon is observed in node error profile curves with respect to the gateway time as a~\textit{dip before steady state}. Ideally, at this point of intersection, labeled as~\textit{Transient Error} in Figure~\ref{fig:errorP}, a node $i$ in the network is perfectly synchronized to the gateway node clock. This is not possible due to the discrete nature of the time estimate update. However, for a certain number of communication cycles, an absolute error well below $\Delta$ can be achieved between the clock of node $i$ and the gateway node clock. A scheme is devised to make nodes recognize and stop at this point. If this task is achieved, synchronization can be achieved at high accuracy using minimal number of communication cycles.

A stopping criterion is therefore needed to halt the iterative process of each node before steady state is reached. The difficulty in developing such a criterion has to do with the node having to rely on the signature of its local time estimate series which the error dip directly causes. This leads to an involved pattern recognition problem. However, the authors in~\cite{al2017efficient} suggested a simple but effective solution for the problem. This solution is in the form of a detection filter which operates on each node's time series to detect the minimum error at the dip region. This filter produces a signal to indicate where in the time series a node's time estimate is closest to the gateway time. To design an optimum filter that halts the time estimate update at the dip requires rigorous analysis which is beyond the scope of this work. However, in~\cite{al2017efficient}, a heuristic finite impulse response filter which exploits the inflection point of the time curve is employed to provide the indicator signal for detecting the dip region. Through experimentation, it was reported that the zero-crossings of the filter output corresponds, with high probability, to the instant of error dip. An example of such a filter was reported in~\cite{al2017efficient} to have an impulse response, $h(k)$ given by:

\begin{equation}\label{eq5}
\begin{aligned}
h(k) ={} & 0.2\delta (k+3) + 0.5\delta (k+2) + 0.2\delta (k+1) \\
& - 0.2\delta (k-3) - 0.5\delta (k-2) - 0.2\delta (k-1)
\end{aligned}
\end{equation}
where $\delta$ is the Khronecker delta function. The above filter is a typical example of a difference filter of length 7.

\subsection{Local Clock Reset and Resynchronization}
The procedure used for local node update has to yield an error dip for all the nodes of the network at almost the same time instant. If the variance in the instances of reaching the dip region for all network nodes is nearly null, then the scheduled wake-up and sleep of the network can be carried out in unison which would make synchronization more energy efficient. Once the time estimate at the transient error is obtained, it is used to reset the physical local clock of the node. After which, a node seizes to update and goes into a sleep mode. The node then awakes after a certain predetermined period when the clocks of the network nodes drift beyond an acceptable threshold. In~\cite{maroti_flooding_2004}, experiments were conducted using mica and mica2 mote platforms to determined the resynchronization period. The authors reported a resynchronization period of $30s$ to give the best results but showed that, when skew compensation is taken into consideration in protocol design, the resynchronization period can go up to several minutes depending on the accuracy requirements of the specific application of the wireless sensor network. In~\cite{karl_protocols_2007}, a simple formula is reported to estimate resynchronization period for a node tat only does offset compensation during synchronization. Given a node with a constant drift rate $x$ given in \textit{parts-per-million ($ppm$)} representing the accuracy of the crystal oscillator of the node clock, if an application desires an accuracy of $\delta \ seconds$, a protocol has to resynchronize every $\frac{\delta}{x \times 10^{-6}} seconds$. Therefore, an oscillator with $100 ppm$ running at 1 MHz drifts apart 100 $\mu$s in one second. For example, if it is desired to have an accuracy of $\delta$ = 1 $ms$ and $x$ =100 ppm, then the resynchronization period must be conducted every 10 seconds.

\section{The Proposed Synchronizer - A General Setup}
\label{The Proposed Synchronizer - A General Setup}
The discrete switched system, which represents the proposed synchronization action of the network as a whole, is realized using two interconnected discrete dynamical systems. The states of the first sub-system ($\textbf{T}_{k}$) describe the time-like dynamical variables used by the nodes to determine the global time of the gateway node. The states of the second dynamical subsystem ($\textbf{J}_{k}$) have a binary nature (0,1) that is used to activate or inhibit a node time series update.  The whole network dynamics is described using the equations in (\ref{eq_asyn}):

\begin{equation}\label{eq_asyn}
\left[ \begin{array}{c} \textbf{T}_{k} \\ \textbf{J}_k \textbf{1}\end{array} \right] = \left[ \begin{array}{c} \textbf{J}_k (\textbf{AT}_{k-1} + \Delta k\textbf{B}) \\ \mathscr{F}(\textbf{J}_{k-1} \textbf{1})\end{array} \right]
\end{equation}

where, $$\textbf{J}_k = \begin{bmatrix}
J_{1k} &  0  & \ldots & 0\\
0  &  J_{2k} & \ldots & 0\\
\vdots & \vdots & \ddots & \vdots\\
0 &  0  & \ldots & J_{Nk}
\end{bmatrix}$$
The non-zero components of $\textbf{J}_k$, at discrete time $k$, are expressed as a function of their state values at the previous time, $k-1$ and the activation function $\mathscr{F}(.)$ is determined by the method that establishes graph connectivity of the WSN. The entries $J_{ik}$ are therefore expressed as:
$$
J_{ik} =
\begin{cases}
1, & \text{if}\ \text{node}\ i\ \text{updates}\  \text{at}\ \text{time} \ k\\
0, & \text{otherwise}
\end{cases}
$$
\\
It is worth noting here that, (\ref{eq_asyn}) becomes a synchronous system  if $J_{ik} = 1 , \ \forall i \in V \ \text{or} \ \textbf{J} = \textbf{I}$, where $\textbf{I}$ is the identity matrix.

The local node time estimate for node $i$ at discrete time $k$ can also be represented as:

\begin{equation}\label{eq_update}
t_i(k) =   \frac{J_{ik}}{|\mathcal{N}_{i}|} \left [\sum_{j \in \mathcal{N}_i}^{} t_j(k-1) \right ]
\end{equation}

where node $i$ is activated for update when $J_{ik}$ is \textquoteleft1\textquoteright \ and inhibited from update when $J_{ik}$ is \textquoteleft0\textquoteright \ and $\mathcal{N}_i$ is the set of nodes that node $i$ communicates with. It is ought to be noted that $\mathcal{N}_i$ may contain the gateway node.

\subsection{Operational Modes of a Sensor Node}
\begin{figure}[ht]
	\centering
	\includegraphics[width=0.4\textwidth]{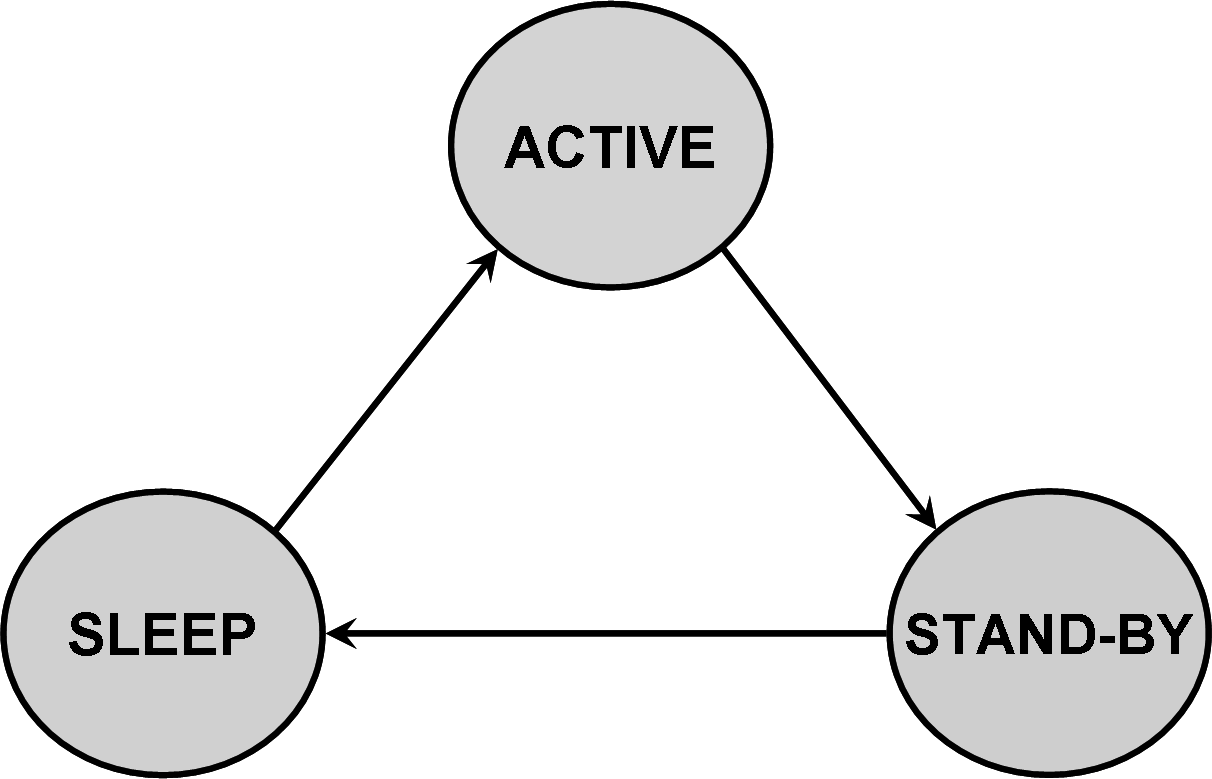}
	\caption[Operational Cycle of a Sensor Node]{Operational Cycle of a Sensor Node}
	\label{fig:modes}
\end{figure}

Due to the limited energy of WSN motes, they are designed to switch between active, sleep and stand-by modes during operations. In \cite{polastre2005telos}, these modes and their energy consumption requirements are presented for Telos, Mica2 and MicaZ mote platforms. The proposed synchronizer exploits the active, sleep and stand-by modes in its asynchronous switching operation in order to conserve energy. The protocol cycles between these modes as illustrated in Figure \ref{fig:modes}.

\noindent
In the active mode, each node can;

\begin{enumerate}
	\item Transmit a request for neighbors time estimates
	\item Transmit it's time estimate
	\item Receive request from neighbors for its own time estimate
	\item Compute and update its time estimate
\end{enumerate}

\noindent
In the stand-by mode, each node can;

\begin{enumerate}
	\item Transmit values of its logical clock
	\item Receive request for its own logical clock
\end{enumerate}	

\noindent
In the sleep mode, each node does;

\begin{enumerate}
	\item Keep physical clock running
	\item Monitor how much time has elapsed in order to switch to active mode
\end{enumerate}

The scenarios, for which the synchronizer is designed, do not permit node communication except with immediate neighbors. These scenarios expect, among other things, that the synchronizer functions normally when extra nodes are added to and/or deleted from the network. This necessitates that the system in~(\ref{eq_update}) be broken-down into identical subsystems cemented together by nearest neighbor communication (i.e. protocols). The nodes of the network may use the protocol collectively to realize the effect of the system in~(\ref{eq_asyn}).

\subsection{Description of the General Synchronizer}
Algorithm \ref{alg1} provides a general description of the protocols while a detailed implementation of three different synchronization realizations are given later in the next section.

\begin{algorithm}
	\caption{Synchronization Pseudo-code for Node $i$}\label{alg1}
	\begin{algorithmic}[1]
		\State \textbf{Initialization}
		\State  Wake-up from sleep mode
		\State  Clear Repository
		\State  $t_c \gets \tau_i(0); t_s \gets 0$;
		\State  Initialize $R$; \Comment{ Resynchronization Period}
		\State  Initialize $\sigma_{max}$; \Comment{ Upper bound on the convergence time}
		\State  $I_{T} \gets 1$; \Comment{Time value request trigger}
		\State  $I_{U} \gets 1$; \Comment{Soft clock estimation trigger}
		\State  $I_{S} \gets 0$; \Comment{Dip stage trigger}
		\State  $I_{R} \gets 0$; \Comment{Response to neighbor request trigger}	
		\If{$I_S = 1$}
		\State $t_i \gets t_s$
		\ElsIf{$I_S = 0$}
		\State $t_i \gets t_c$
		\EndIf
		
		\While{$I_U = 1$}
		\If{$I_T = 1$}	
		\State Send request for $<t_j>$ from nodes $j \in \mathcal{N}_i$
		\EndIf	
		\State \textbf{Upon receiving all $<t_j>$s}
		\State Compute neighborhood time estimate, $t_s$
		\For{Each estimate $t_s$}	
		\If{$t_s$ is at dip stage}
		\State Set $I_S \gets 1$
		\State Set $t_c \gets t_s$
		\ElsIf{$t_s$ is not at dip stage}
		\State Set $I_S \gets 0$
		\EndIf				
		\EndFor	
		\EndWhile
		\While{$I_S = 1$}
		\State 	Set $I_U \gets 0$ and $I_T \gets 0$
		\State Switch node to stand-by mode
		\State Set timer-1 to fire after $\sigma_{max}$ seconds
		\State Until timer-1 timeout, \If{a request is received from node $j$} \State Set $I_R \gets 1$ \ElsIf{No request is received} \State $I_R \gets 0$ \EndIf
		\EndWhile
		
		\If{$I_S = 1$ and $I_R = 0$}
		\State Set timer-2 to fire after $R$ seconds
		\State Node goes to sleep mode \EndIf
		\If{$I_R = 1$}
		\State Send $<t_i>$ and Reset timer-1 after timeout
		\EndIf
		\State \textbf{Upon timer-2 timeout}
		\State Node wakes up to resynchronize
	\end{algorithmic}
\end{algorithm}

\begin{figure*}[ht!]
	\centering
	\includegraphics[width=1.0\textwidth]{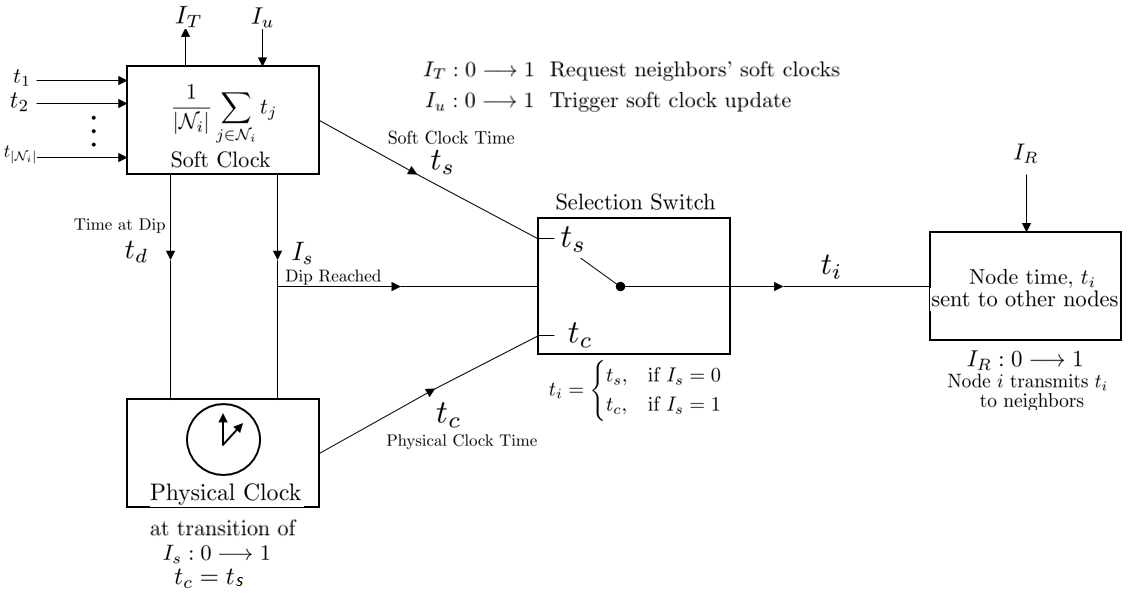}
	\caption[General Synchronizer Block Diagram]{General Synchronizer Block Diagram}
	\label{fig:general_sync}
\end{figure*}

The general realization for the suggested time synchronization protocol is illustrated by Figure~\ref{fig:general_sync} for an activated node $i$ having $\mathcal{N}_{i}$ nearest neighbors and deployed in a network of size $N$. Node $i$ maintains two variables related to its clock(Line 4). The first variable is a time soft clock variable, $t_s$ which is an estimation of the gateway clock representing the global time estimate. The second is a logical clock variable, $t_c$ which is a representation of the physical clock of node $i$ and related to the hardware clock, $\tau_i$. It also maintains four binary variables (Lines 7-10) used to trigger specific actions. These variables are described as follows:

\begin{itemize}
\item  $I_{T}$ triggers when a node requests for neighbors' time values and initialized to \textquoteleft1\textquoteright
\item  $I_{U}$ triggers when the node updates its time estimate initialized to \textquoteleft1\textquoteright
\item  $I_{S}$ is a dip stage detection trigger variable initialized to \textquoteleft0\textquoteright
\item  $I_{R}$ triggers response to neighbor request for node time initialized to \textquoteleft0\textquoteright
\end{itemize}

The following describes how the protocol operates. Node $i$ initializes $t_c$ to the current hardware clock value, $\tau_i(0)$ and set its beaconing rate, $R$ to a predetermined resynchronization period  \footnote{Since the drift rates of WSN nodes' clocks are all time varying, they drift apart in synchronization with time and hence a periodic reinitialization of the protocol is needed after a certain time of drift}. A variable $\sigma_{max} \lll R $ representing an upper bound on the variance in the convergence time between all network nodes is also initialized. Node $i$ then sends requests for neighborhood estimates (Line 17). While $I_{U} = 1$ (update trigger is on), if $I_{T} = 1$ (time request trigger is on), node $i$ sends request for neighborhood time values. If nodes $j, \ j \in \mathcal{N}_i$ receive request from node $i$, each node $j$ replies with an acknowledgment having  its current time value, $t_j$ as payload. With 1-hop communication, node $i$ receives all $t_j$ at approximately the same time. Upon receiving all $t_j$, node $i$ calculates its current time estimate, $t_s$ using a local time update algorithm (Line 19). For each estimate $t_s$, node $i$ checks if $t_s$ is in the dip region, if true, it sets $I_{S} = 1$ and $t_{c} = t_s$, if not, $I_{S}$ is maintained at $0$ (Lines 20-25). As shown in Figure \ref{fig:general_sync}, $t_i$ takes the value of $t_c$ when $I_{S} = 0$ but is set to $t_s$ when $I_{S} = 1$, which is akin to a selection switch with $I_{S}$ as the decision variable (Lines 11 to 14). While $I_{S} = 1$, $I_{U}$ and $I_{T}$ are set to \textquoteleft0\textquoteright, the node switches to the standby mode, and sets a timer-1 to fire after $\sigma_{max}$ seconds. Where the update trigger variable $I_U$ is equivalent to $J_{ik}$ in (\ref{eq_update}). The time request trigger, $I_{R}$ is set to \textquoteleft1\textquoteright, when timer-1 times out (Lines 26-30). If $I_{S} = 1$ and $I_{R} = 0$ (i.e., no request is made from neighbors and the node $i$ is in standby mode), a timer-2 is set to fire after $R$ seconds and the node goes to sleep mode(Lines 35-37). $I_{R}$ is set to \textquoteleft1\textquoteright \ anytime a request is received from neighbors and set to \textquoteleft0\textquoteright \ otherwise (Lines 31-34). If $I_{R} = 1$, node $i$ transmits its time $t_i$ to its neighbors and resets timer-1 after its times out. The node remains in the sleep mode until timer-2 times out, when it awakes to resynchronize.

\section{Realizations of the Proposed Asynchronous Paradigm}
\label{Realizations of the Proposed Asynchronous Paradigm}
In this section, three realizations of the proposed synchronization protocol are presented. The pseudo-code and  description of each realization aided by relevant diagrams are provided. It is assumed that for all methods, the gateway node ticks uniformly at a rate $\Delta$, represented by $t_g(k)=\Delta \times k$. This assumption is practical since many sensor network architectures consist of a gateway or base-station that has access to a stable and precise clock reference such as in a Global Positioning System (GPS) receiver \cite{Lenzen2015,akl_investigation_2011}. In the absence of an external clock source, the hardware clock of a selected sensor node, e.g. the gateway(root) node, might serve as a source. All the three realization are shown here for a 9 node grid topology network. The application of these realizations to any other type or size network is straight forward.

\subsection{Timed Sequential Asynchronous Update (TSAU)}
\begin{figure}[ht]
	\centering
	\includegraphics[width=0.485\textwidth]{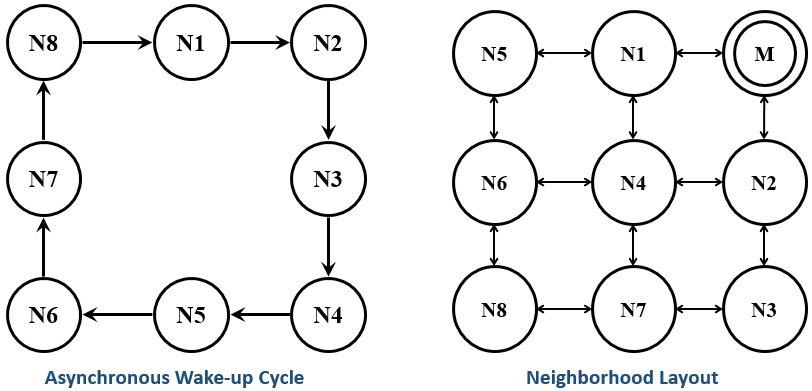}
	\caption[Asynchronous Wake-up Cycle and Operation Mechanism for TSAU]{Asynchronous Wake-up Cycle and Operation Mechanism for TSAU}
	\label{fig:tsap}
\end{figure}

In the first realization named Timed Sequential Asynchronous Update (TSAU), network nodes are made to compute time estimates one at a time asynchronously in a sequential manner. The sequence is based on proximity to the gateway node, i.e. the closer a node is to the gateway node, the earlier it updates. This is illustrated by Figure ~\ref{fig:tsap}. The pseudo code of TSAU is given by \textbf{Algorithm 2}.

{Operation Mechanism}:
Initially when the node is powered on, two variables, $clockSum$ and $numReceived$ required to calculate the average synchronization time estimate to the neighboring nodes are initialized to zero.
Whenever a synchronization message from any neighboring node is received, the time estimate $t_j$ of a neighboring node $j$ at the time of reception, is saved. The received time, $t_j$ is then added to the $clockSum$ variable and the number of received clock values, $totalReceived$ is incremented. To exchange time synchronization packets with its neighboring nodes, node $i$ transmits a broadcast packet of its current time information approximately every $\Delta$ seconds. In every $\Delta$ seconds, when the number of received clock values is more than one, the time variable of node $i$, $t_i$ is updated by setting it to the average estimate, $t_{av}$. Finally, the time estimate of node $i$, $t_i$ is  transmitted and the $clockSum$ variable and the number of received clock values, $totalReceived$ are initialized.

In order for nodes to update in an atomic manner, node IDs are assigned based on proximity to the gateway node. We program each node to update when the variable $UpdateTime$ is a multiple of $\Delta$. This variable is calculated by $UpdateTime += (N-1) \times \Delta$ where $N$ is the number of network nodes including the gateway node. The $UpdateTime$ for a node with ID, $\textit{i}$ is initialized to $\textit{i} \times \Delta$. This allows the activation of the nodes to be carried out one at a time in a sequential manner.

\begin{table}[htbp]
	\centering
	\begin{tabular}{lllllll}
		\toprule
		\multicolumn{7}{c}{\textbf{Algorithm 2: TSAU Pseudo-Code for Node $i$}} \\
		\midrule
		\multicolumn{7}{c}{} \\
		\multicolumn{7}{l}{\textbf{Initialization}} \\
		\multicolumn{7}{c}{$t_i\longleftarrow t_i(0)$;
			$UpdateTime \longleftarrow \textit{i} \times \Delta$;}\\
		\multicolumn{7}{c}{$clockSum\longleftarrow 0$; $totalReceived\longleftarrow 0$;}\\
		\multicolumn{7}{l}{\textbf{If $<UpdateTime>$ is a multiple $\Delta $}} \\		
		\multicolumn{7}{l}{\textbf{If $<t_j>$ is received}} \\
		\multicolumn{7}{l}{$clockSum\longleftarrow clockSum + (t_j)$;} \\
		\multicolumn{7}{l}{$totalReceived\longleftarrow totalReceived + 1$;} \\
		\multicolumn{7}{l}{$t_{av}\longleftarrow clockSum/totalReceived$;} \\
		\multicolumn{7}{l}{$UpdateTime\longleftarrow UpdateTime + (N-1) \times \Delta$;} \\		
		\multicolumn{7}{c}{} \\
		\multicolumn{7}{l}{\textbf{Upon every $\Delta$ seconds}} \\
		\multicolumn{7}{l}{If $totalReceived > 1$ and $IncomingID = Node j$} \\
		\multicolumn{7}{l}{$t_i \longleftarrow  t_{av}$ } \\
		\multicolumn{7}{c}{} \\
		\multicolumn{7}{l}{ $clockSum\longleftarrow 0$; $totalReceived\longleftarrow 0$;} \\
		\multicolumn{7}{l}{broadcast $<t_i>$} \\
		\bottomrule
	\end{tabular}%
	\label{tab:addlabel}%
\end{table}%

\noindent
$\square$ \ \textbf{Remark 1}:
It should be noted that this realization operates in a completely blind fashion since it neither requires to know the sender node nor store its time information. Therefore the protocol is expected to show some robustness to changes in network topology. However, for this protocol to work properly, the network topology has to be maintained throughout its operation. If deployed in a network whose graph is time variant, a wake-up activation sequence which might require different communication packet(s) will be required to control the sequential update of the nodes. In that case, a fixed network topology is not necessary but full knowledge of the network nodes and connectivity are needed.

\subsection{Unidirectional Asynchronous Flooding (UAF)}
In the second realization, named Unidirectional Asynchronous Flooding (UAF), nodes update asynchronously based on a wake-up activation protocol regulated by the gateway node. This method improves upon TSAU, whereby it is designed to make nodes at approximately the same proximity to the gateway node update at the same communication instant. This is based on the assumption that the transmission and reception times of messages to and from their neighbors and/or to the gateway node are the same. The gateway node is made to regulate the activation of nodes by flooding the network with wake-up messages in every cycle of asynchronous update to begin the cycle. The cycle is re-initiated  once a current cycle completes, hence the name \textit{Unidirectional}.

\begin{figure}[ht]
	\centering
	\includegraphics[width=0.50\textwidth]{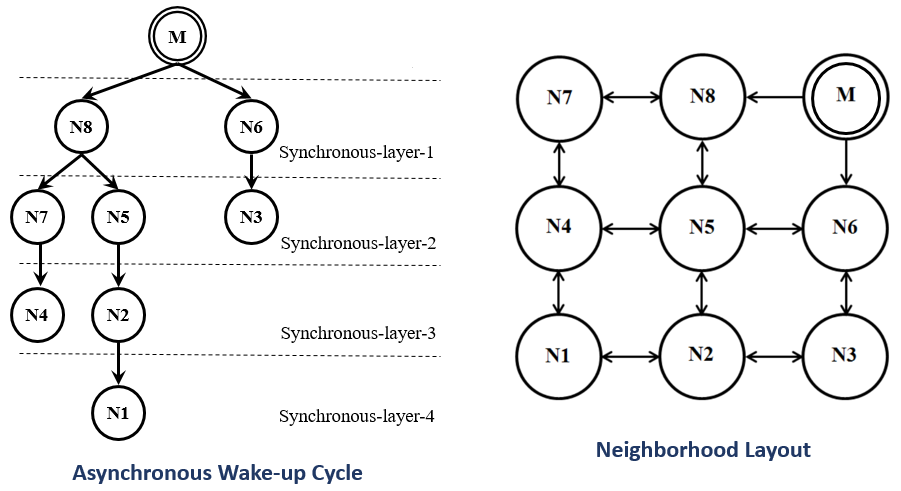}
	\caption[Asynchronous Wake-up Cycle and Operation Mechanism for UAF]{Asynchronous Wake-up Cycle and Operation Mechanism for UAF}
	\label{fig:tauf}
\end{figure}

{Operation Mechanism}:
The description of this asynchronous activation cycle procedure is illustrated by Figure ~\ref{fig:tauf}. As shown in Figure ~\ref{fig:tauf}, layers of connectivity are defined for all network nodes based on their proximity to the gateway node.

All nodes belonging to the same connectivity layer wake up at the same time to carry out the averaging process, hence the name \textit{Synchronous Layer}.  To elaborate further on the operation of the method, we outline its stages of operation in the pseudo-code of \textbf{Algorithm 3} which are described as follows:

\begin{enumerate}
	\item Each node, $i$ has a binary status variable labeled $s_i$ that is set to, $s_i=0$. Let us assume an upper bound $L$ on the number of connectivity layers, where $L$ depends on the network size and topology. For example, for the network in Figure ~\ref{fig:tauf}, $L = 4$.
	\item The gateway node initializes update timer $T_s$ and triggers the update of the nodes connected to it.
	\item Once a node $i$ updates, it triggers the update of its nearest neighbor nodes, $j$ whose status bit variable, $s_j$ are a complement of its own, i.e., $s_i=\acute{s}_{j}$. Once the flooding of the status bits variable begin, if $i$ receives  $<t_j,s_j>$ such that, $s_i=\acute{s}_{j}$ then node $i$ accepts the clock value of $j$ and computes its average, then updates its clock with the computed average and set its $s_i$ to $s_j$. If on the other hand, $s_i = s_j$, then it means that nodes $i$ and $j$ belong to the same connectivity layer or the connectivity layer of $j$ lies above  that of $i$ therefore it doesn't wake-up to compute and update its clock and hence conserve energy.
	\item Each node then broadcasts whatever values of $<t_i,s_i>$ it has every $\Delta$ seconds.
	\item This process continues until the timer of the gateway node is $Ts > L \times \Delta$. When this event is true, the gateway node initializes its update time $T_s$ and trigger the update of the nearest nodes and hence the whole process begins again.
\end{enumerate}

\begin{table}[htbp]
	\centering
	\begin{tabular}{lllllll}
		\toprule
		\multicolumn{7}{c}{\textbf{Algorithm 3: \textbf{UAF} Pseudo-Code for Node $i$}} \\
		\midrule
		\multicolumn{7}{c}{} \\
		\multicolumn{7}{l}{$\square$ \textbf{Initialization}} \\
		\multicolumn{7}{l}{$t_i\longleftarrow t_i(0)$; $s_i \longleftarrow 0$;
			$clockSum \longleftarrow 0$; $totalReceived\longleftarrow 0$;}\\
		\multicolumn{7}{l}{$\square \ \textbf{Upon \ receiving} <t_j,s_j>$} \\		
		\multicolumn{7}{l}{\textbf{If} $<s_j \neq s_i>$ \textbf{then}} \\
		\multicolumn{7}{l}{$clockSum\longleftarrow clockSum + (t_j)$} \\
		\multicolumn{7}{l}{$totalReceived\longleftarrow totalReceived + 1$} \\
		\multicolumn{7}{l}{$t_{av}\longleftarrow clockSum/totalReceived$} \\
		\multicolumn{7}{l}{$s_i \longleftarrow  s_j$ } \\		
		\multicolumn{7}{l}{\textbf{else if} $<s_j = s_i>$ \textbf{then} $t_{av}\longleftarrow t_i$ \textbf{endif}}\\
		\multicolumn{7}{c}{} \\
		\multicolumn{7}{l}{$\square$ \textbf{Upon every $\Delta$ seconds}} \\
		\multicolumn{7}{l}{$t_i \longleftarrow  t_{av}$ } \\
		\multicolumn{7}{c}{} \\
		\multicolumn{7}{l}{ $clockSum\longleftarrow 0$; $totalReceived\longleftarrow 0$;} \\
		\multicolumn{7}{l}{broadcast $<t_i,s_i>$} \\
		\bottomrule
	\end{tabular}%
	\label{tab:alg1}%
\end{table}%

\noindent
$\square$ \ \textbf{Remark 2}:
It should be noted here that, the only knowledge needed for this method is a loose upper bound on the connectivity layers at the gateway node. Also, no knowledge of network size and/or constituents is needed.

\subsection{Bidirectional Asynchronous Flooding (BAF)}
In this section we present another realization of the synchronization scheme. This realization does not require any regulation of the update wake-up sequence by the gateway node. Hence it can effectively operate in a sparsely distributed random network. In this method, all variables are similar to those in \textbf{Algorithm 3} except another variable is introduced to make the network self-regulating in carrying out the update wake-up cycle and therefore removes the need for the centralized wake-up regulation seen in UAF. To achieve this, once the update wake-up cycle is initiated by the gateway node, the node(s) in the first and the last connectivity layers automatically carry out the regulation and does not depend on the gateway node for the regulation. There are two realizations of connectivity layers: a Forward Synchronous Layer and a Backward Synchronous Layer.

\begin{figure*}[ht]
	\centering
	\includegraphics[width=0.90\textwidth]{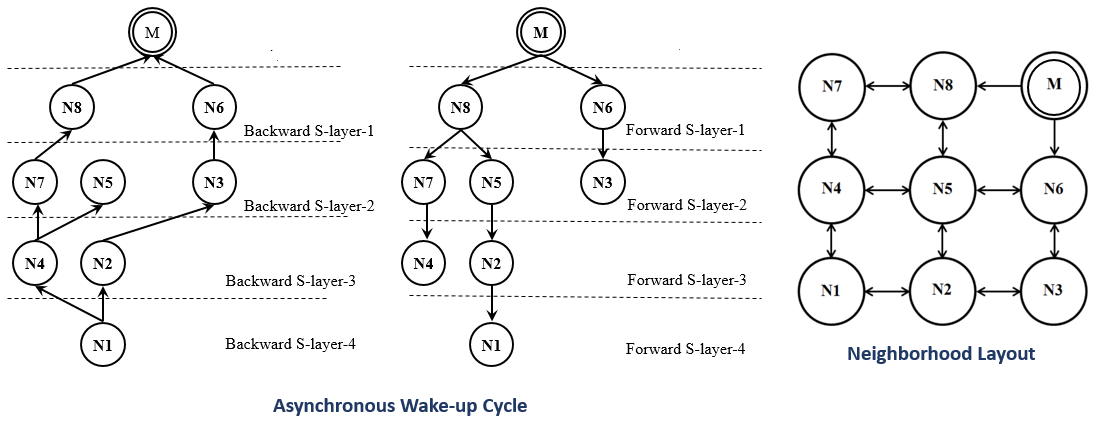}
	\caption[Asynchronous Wake-up Cycle and Operation Mechanism for BAF]{Asynchronous Wake-up Cycle and Operation Mechanism for BAF}
	\label{fig:bafp}
\end{figure*}

{Operation Mechanism}:
The pseudo-code is presented in \textbf{Algorithm 4} and described as follows;
\begin{enumerate}
	\item Each node, $i$ has a binary status variable labeled as $s_i$ that is set to, $s_i=0$ and also has a counter variable $c_i$ that is also initially set to zero.
	\item The gateway node triggers the update of the nodes by broadcasting its clock values.
	\item Once any node receives the gateway clock value (i.e. nodes in the Forward-S Layer-1 as shown in Figure ~\ref{fig:bafp}), the node updates its clock and negates its status bit variable.
	\item After doing stage 3, the node triggers the update of its nearest neighboring nodes whose status bit is the complement of its own.
	\item When a node $i$ receives a packet from another node $j$, it compares its $s_i$ with the received $s_j$ and if $s_i\neq s_j$ it means node $j$ belongs to a layer that triggers the update of node $i$'s layer.
	\item If the event in 4 is true, then node $i$ accepts the time estimate of $j$ and computes its average, then save its computed average time estimate and set it's $s_i$ to $s_j$ and also set its $c_i$ to $c_i+1$. If on the other hand, if $s_i = s_j$, then it means that nodes $i$ and $j$ belong to the same connectivity layer or the connectivity layer of $j$ does to trigger the update of $i$ for the current wake-up cycle and therefore it doesn't wake-up to compute and update its clock and hence conserves energy.
	\item The process continues until a node finds that its $c_i$ variable is the highest compared to the connected node and their $s_i$'s are the same, then this is the furthest node from the gateway node.
	\item This furthest node sets its $c_i = 0$, negate its $s_i$ and trigger a backward flooding.
	\item In the backward flooding, all network nodes now have the complement of their initial $s_i$'s , i.e. $\acute{s}_i$ hence the process automatically continues until the first layer is reached, which in-turn trigger the next forward flooding.
\end{enumerate}

\begin{table}[htbp!]
	\centering
	\begin{tabular}{lllllll}
		\toprule
		\multicolumn{7}{c}{\textbf{Algorithm 4: \textbf{BAF} Pseudo-Code for Node $i$}} \\
		\midrule
		\multicolumn{7}{c}{} \\
		\multicolumn{7}{l}{$\square$ \textbf{Initialization}} \\
		\multicolumn{7}{l}{$t_i\longleftarrow t_i(0)$; $s_i \longleftarrow 0$;$c_i \longleftarrow 0$;$clockSum \longleftarrow 0$; $totalReceived\longleftarrow 0$;}\\
		\multicolumn{7}{l}{$\square \textbf{Upon \ receiving} <t_j,s_j,c_j>$} \\		
		\multicolumn{7}{l}{\textbf{If} $<s_j \neq s_i>$ \textbf{then}} \\
		\multicolumn{7}{l}{$clockSum\longleftarrow clockSum + (t_j)$} \\
		\multicolumn{7}{l}{$totalReceived\longleftarrow totalReceived + 1$} \\
		\multicolumn{7}{l}{$t_{av}\longleftarrow clockSum/totalReceived$} \\
		\multicolumn{7}{l}{$s_i \longleftarrow  s_j$ } \\		
		\multicolumn{7}{l}{$c_i \longleftarrow  c_j + 1$ } \\		
		\multicolumn{7}{l}{\textbf{else if} $<s_j = s_i>$ \textbf{then} $t_{av}\longleftarrow t_i$ \textbf{endif}}\\
		\multicolumn{7}{l}{\textbf{if} $<s_j = s_i> \ \textbf{and} \ c_i > c_j, \forall j  $ \textbf{then} $c_i \longleftarrow 0 \ and \ s_i \longleftarrow \bar{s}_i$ \textbf{endif}}\\
		
		\multicolumn{7}{c}{} \\
		\multicolumn{7}{l}{$\square$ \textbf{Upon every $\Delta$ seconds}} \\
		\multicolumn{7}{l}{$t_i \longleftarrow  t_{av}$ } \\
		\multicolumn{7}{c}{} \\
		\multicolumn{7}{l}{ $clockSum\longleftarrow 0$; $totalReceived\longleftarrow 0$} \\
		\multicolumn{7}{l}{broadcast $<t_i,s_i,c_i>$} \\
		\bottomrule
	\end{tabular}%
	\label{tab:addlabel}%
\end{table}%

\noindent
$\square$ \ \textbf{Remark 3}:
As observed from the operation of BAF, once the gateway node triggers the initial asynchronous process, the network nodes automatically continues to carry-out the update wake-up cycles whiles making use of the proximity to the gateway node, asynchronicity and energy conservation from controlled and limited computations. This feature make this version of the proposed scheme very robust in that, even if the connectivity layers of nodes are not maintained with time due to node failure and changes in network topology, new layers of connectivity are automatically formed based on the capacity of the nodes in terms of transmission range and the update wake-up cycle continues to be regulated. The only added requirement here is the increased bytes introduced by the added counter variable, $c_j$.

\section{Simulation Evaluation}
\label{Simulation Evaluation}
To evaluate the performance of the suggested methods for node time update, simulations are carried out on MathWorks MATLAB platform. In each simulation, it is assumed that there are perfect network conditions with no delays, jitters, fading and noise. Each node time is initialized as a random scalar between 0 and 1, i.e. $t_i(k=0) = rand[0-1]$ and the gateway time is initialized to 0, i.e. $t_g(k=0) = 0$. The incremental time step and ticking rate of the gateway node $\Delta$ is taken to be $1 ms$. Each node time estimate is carried out iteratively by taking the average of its neighborhood time values. The simulation parameters are given in Table \ref{table:: Parameter Specifications for Matlab Simulation}.

\begin{table}[htbp!]
	\caption{Parameter Specifications for MATALB Simulation}
	\centering
	\label{table:: Parameter Specifications for Matlab Simulation}
	\begin{tabular}{ l  c  c }
		\toprule
		Parameter & Value & Unit \\
		\midrule
		Incremental Time-Step Size, $\Delta$ & 0.001 & sec \\
		\midrule
		Initial slave time, $t_i(k=0)$ & rand[0.0-1.0] & sec \\
		\midrule
		Initial master time, $t_g(k=0)$ & 0 & sec \\
		\bottomrule
	\end{tabular}
\end{table}

\subsection{Performance Metrics for Evaluation}
To compare the performance of the developed schemes, three parameters are employed. First the minimum time error of each node in the dip region is compared to the gateway node time, followed by the number of iterations needed to reach that minimum error. Finally the variance in the number of communication cycles needed to reach the dip region is compared for all network nodes. Each metric is described below.
\subsubsection{Minimum Error in Dip Region}
The minimum time error in the dip region gives a sense of how accurate a protocol is. An effective method is expected to give very small error values in the dip region. This error is determined by comparing the time values of each node to the gateway node and taking the minimum. For $N$ network nodes, this can be measured by taking the average of the minimum error in the dip region of all the nodes and represents the expectation in the minimum error.

$$\mathbb{E}^{dip}_{min} = \frac{1}{N-1} \sum^{}_{i \in V, i \neq g} \min\limits_{i\in V, i \neq g} e_i(k)$$.

Where $e_i(k) = |t_g(k) - t_i(k)|$

\subsubsection{Communication Cycles to Reach Dip Region}
This parameter is obtained by looking at the number of communication cycles at which the minimum error in the dip region occurs and can be measured by taking the average of this value for all the nodes and is denoted, $k^{dip}_{min}$ and can be defined as:

$$
k^{dip}_{min} = \frac{1}{N-1} \sum^{}_{i \in V, i \neq g} k^{dip}_{i}
$$

Where $k^{dip}_{i} = \{ k:e_i(k) = \min\limits_{i\in V, i \neq g} e_i(k), \forall k \}$

The lower this parameter is, the lower the energy needed by a protocol to reach the dip region is and hence the more efficient the protocol is.

\subsubsection{Variance in Communication Cycles to Reach Dip Region}
It is more efficient also to have all nodes go to the sleep mode at the same time instant. If this is achieved, the local clock reset and sleep for all nodes can occur at same time instant and hence nearly all nodes will synchronize with the gateway node using approximately the same number of pooling cycles. This parameter is calculated by taking the variance in the number of communication cycles needed to reach the minimum error, $k^{dip}_{i}$ in the dip region for all network nodes. This parameter is denoted, $\mathbb{V}_{k^{dip}_{min}}$ and can be written as:

$$
\mathbb{V}_{k^{dip}_{min}} = \frac{1}{N-1} \sum^{}_{i \in V, i \neq g} ( k^{dip}_{i} - E[k^{dip}_{i}] ) ^2
$$

Where $E[.]$ is the expectation.

\subsection{Standard Cases: Error Profile and Node Times against Gateway Time}

\begin{figure}[ht]
	\centering
	\includegraphics[width=0.48\textwidth]{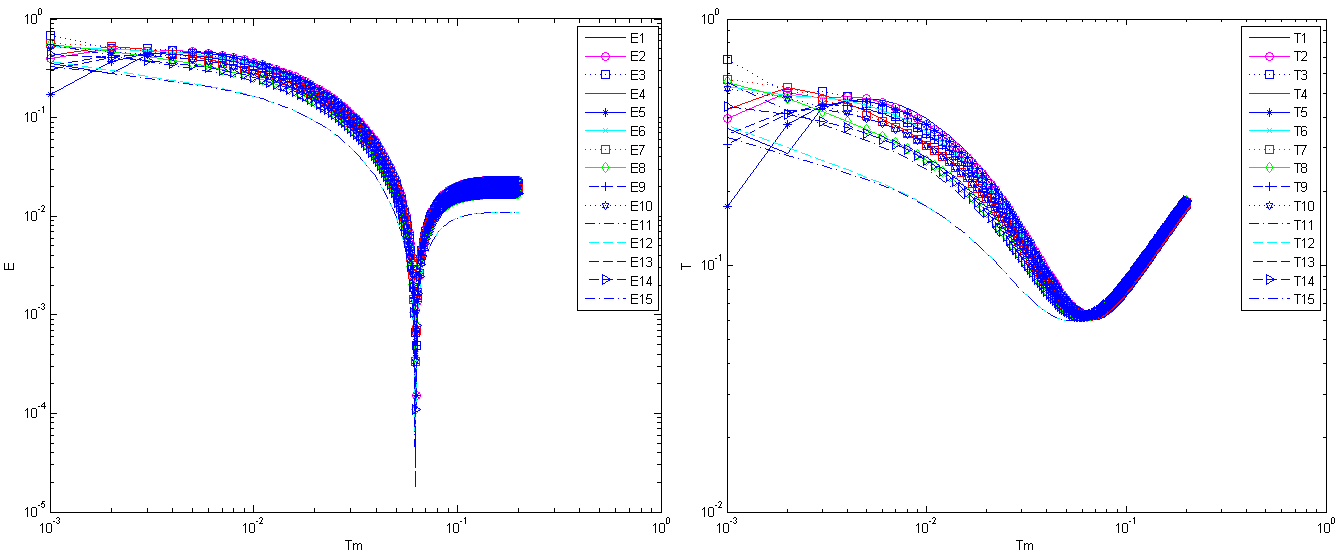}
	\caption[Node Time and Error Profiles for 16 Node Grid Topology for TSAU]{Node Time and Error Profiles for 16 Node Grid Topology for TSAU}
	\label{fig:16Grid}
\end{figure}

\begin{figure}[ht]
	\centering
	\includegraphics[width=0.48\textwidth]{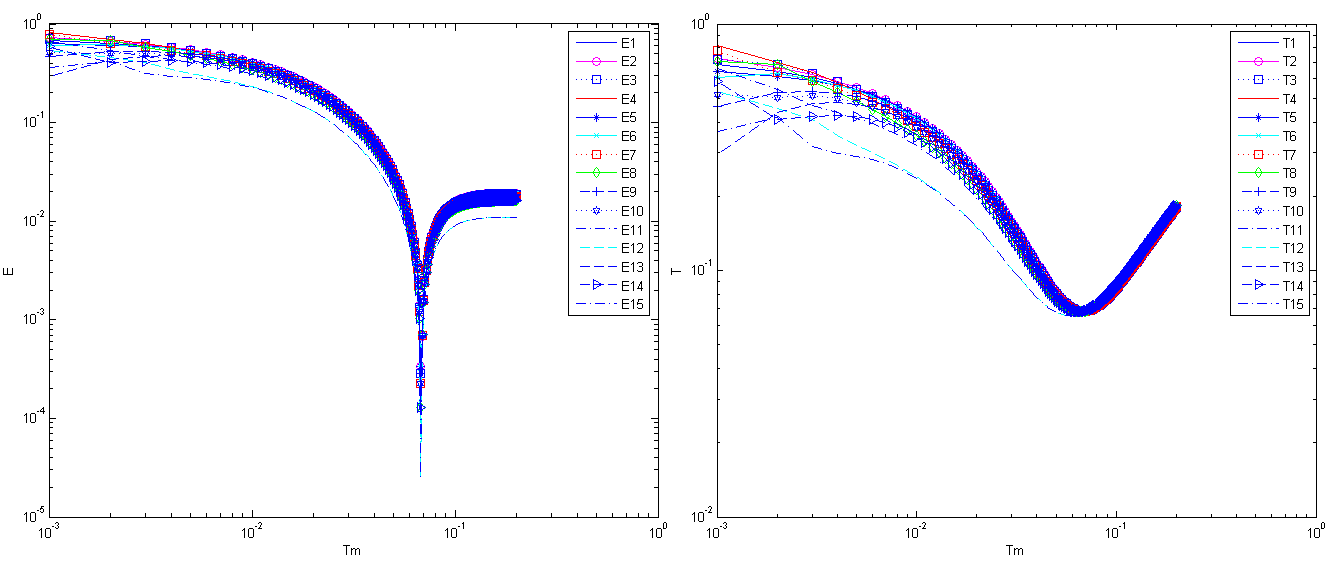}
	\caption[Node Time and Error Profiles for 16 Node Grid Topology for UAF]{Node Time and Error Profiles for 16 Node Grid Topology for UAF}
	\label{fig:16GridT}
\end{figure}

\begin{figure}[ht]
	\centering
	\includegraphics[width=0.48\textwidth]{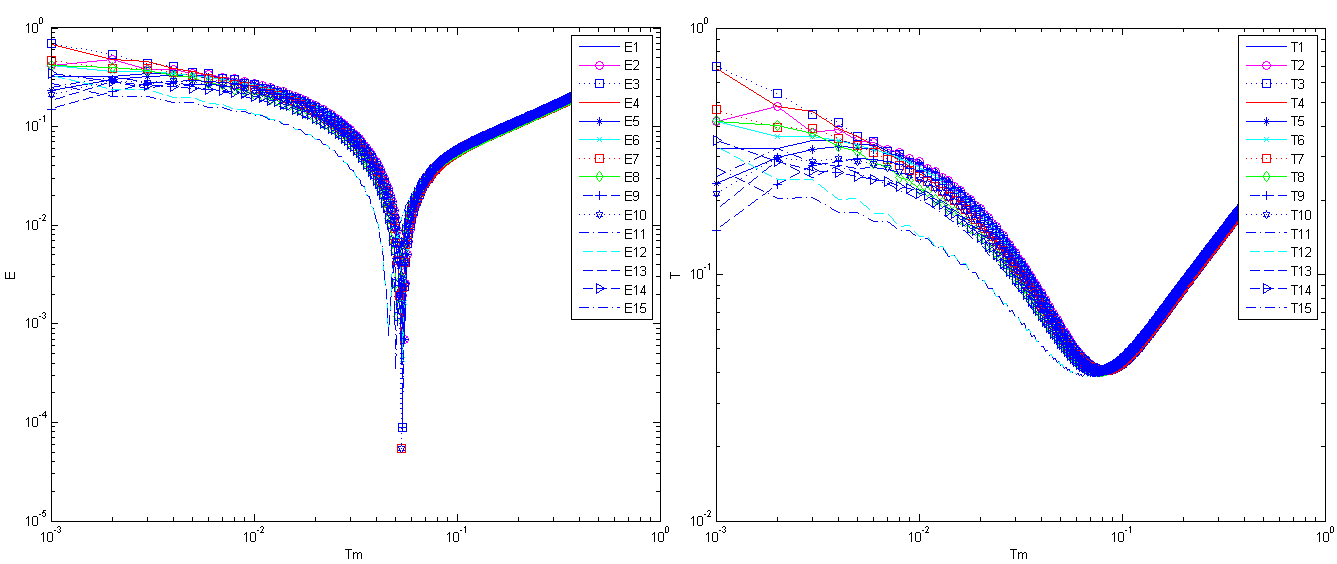}
	\caption[Node Time and Error Profiles for 16 Node Grid Topology for BAF]{Node Time and Error Profiles for 16 Node Grid Topology for BAF}
	\label{fig:16GridB}
\end{figure}

Figures \ref{fig:16Grid}, \ref{fig:16GridT}, and \ref{fig:16GridB} show the error profile plots and node time plots both against the gateway time of a 16 node grid network given in Figure \ref{fig:topps}a for TSAU, UAF and BAF respectively.

We observe from these figures that the dip phenomenon is presented in all nodes using the three methods in the 16 node grid network with an average minimum error $\mathbb{E}^{dip}_{min}$ in the $10^{-4}$ range. It is further observed that BAF registered the lowest error as compared to the other methods. This is followed by UAF, then TSAU.
For $k^{dip}_{min}$ metric, BAF outperforms UAF by a margin of about 16 iterations and TSAP outperforms UAF by a margin of about 6 iterations.

For $\mathbb{V}_{k^{dip}_{min}}$ metric, it is observed that UAF performs the best by having zero variance for the 16 node network. This is a crucial feature since all nodes can stop update, sleep, wakeup and activate at the same instance in time. This unique feature, if manifested in the experimental results, would give UAF a clear advantage over all the other methods. TSAP has the next best performance, followed by BAF.

\subsection{Simulation of Network with Stochastic Link Connectivity}
\begin{figure}[ht]
	\centering
	\captionsetup{justification=centering,margin=2cm}
	\includegraphics[width=0.4\textwidth]{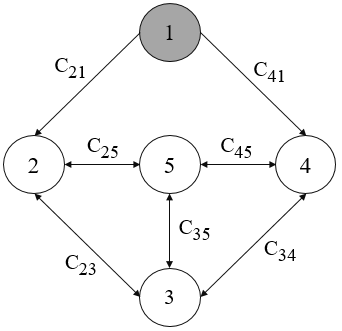}
	\caption[Network for Channel Availability Test]{Network for Channel Availability Test}
	\label{fig:randNet}
\end{figure}

To test the error profile created by the averaging procedure, in particular the transient dip in the error profile \footnote{Refer \cite{al2017efficient} Section II, C}, the test network in Figure \ref{fig:randNet} is simulated on a network with random connectivity links. Node 1 is selected as the gateway node and a communication rate of 1 KHz is used. A connectivity function, $f(\text{C}_{ij})$ among nodes is modeled as a Bernoulli random variable as shown in Figure \ref{fig:randNet} with a discrete probability distribution function given by:
\begin{equation}\label{eq_ben}
f(\text{C}_{ij}) =
\begin{cases}
p & \text{if}\ \text{C}_{ij} = 1\\
1-p & \text{if}\ \text{C}_{ij} = 0\\
\end{cases}
\end{equation}

where $p$ is the probability that the channel is on and the nodes are communicating, and $\text{C}_{ij}$ is the channel connectivity binary variable where $'1'$ indicates that the channel between nodes $i$ and $j$ is active and functional and $'0'$ otherwise.

\begin{figure*}[ht!]
	\centering
	\includegraphics[width=0.9\textwidth]{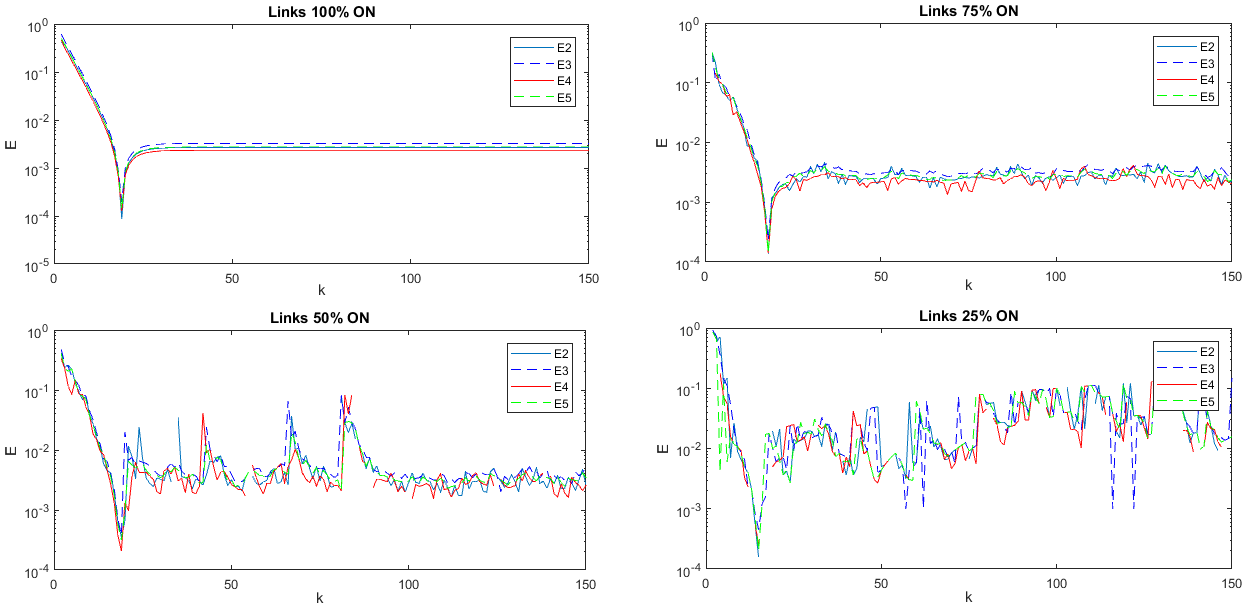}
	\caption[Node Error profile for the Test Network for Different Probabilities of Channel Availability for TSAU]{Node Error profile for the Test Network for Different Probabilities of Channel Availability for TSAU}
	\label{fig:links_tsap}
\end{figure*}

\begin{figure*}[ht!]
	\centering
	\includegraphics[width=0.9\textwidth]{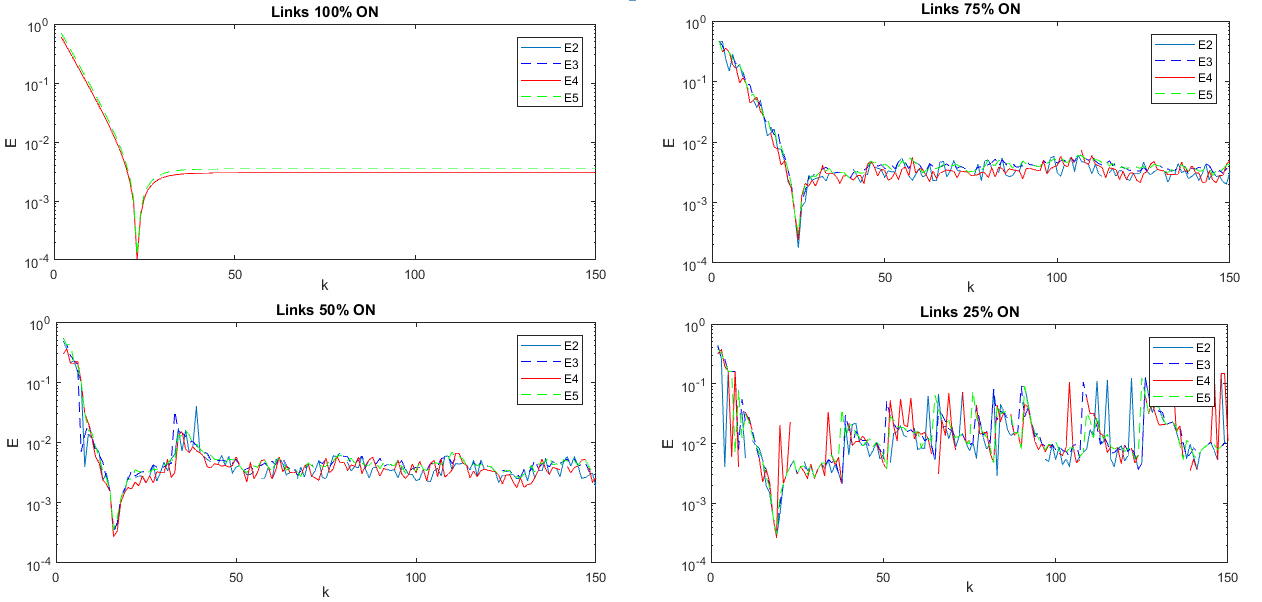}
	\caption[Node Error profile for the Test Network for Different Probabilities of Channel Availability for UAF]{Node Error profile for the Test Network for Different Probabilities of Channel Availability for UAF}
	\label{fig:links_uaf}
\end{figure*}

\begin{figure*}[ht!]
	\centering
	\includegraphics[width=0.90\textwidth]{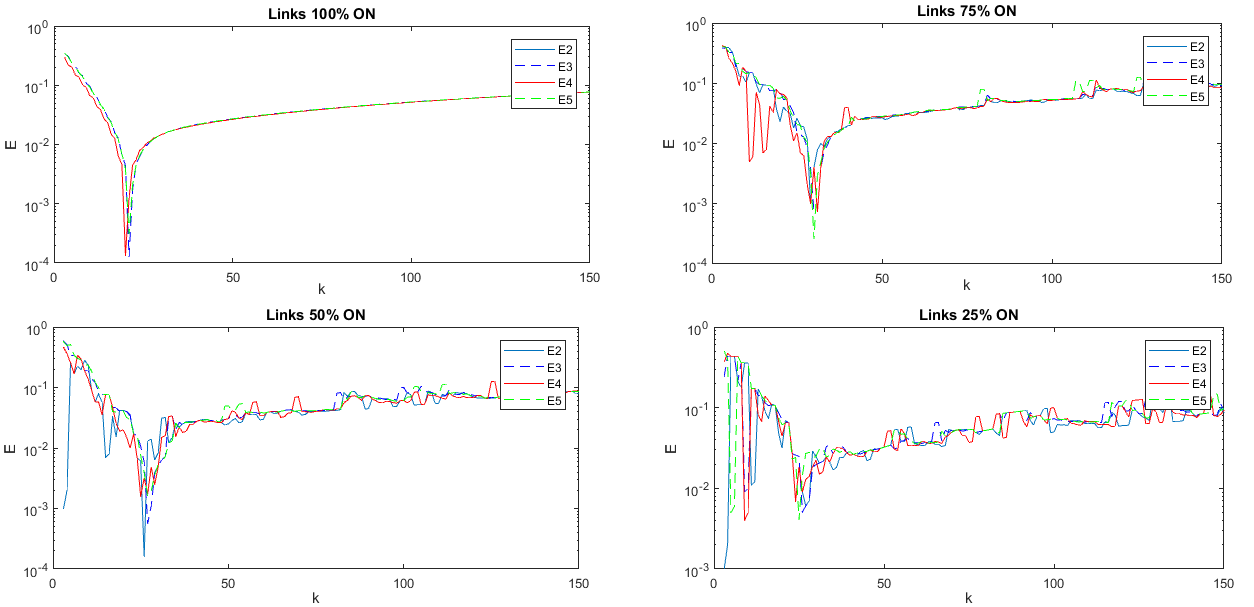}
	\caption[Node Error profile for the Test Network for Different Probabilities of Channel Availability for BAF]{Node Error profile for the Test Network for Different Probabilities of Channel Availability for BAF}
	\label{fig:links_baf}
\end{figure*}

Figures \ref{fig:links_tsap}, \ref{fig:links_uaf} and \ref{fig:links_baf} show the error profiles for TSAU, UAF and BAF respectively. The error profile represents a graph of the error between the evolving time series of each node and the gateway node. For each method, we plot the error profile for each node taken at four values for $p$: $p=1$ where the links are available 100\% of the time, $p=0.75$, $p=0.5$ and $p=0.25$ where communication links malfunction 25\%, 50\% and 75\% of the time, respectively.

It can be seen from these figures that the dip characteristic persists in each profile at all levels of connectivity for all methods. As expected, the error profile becomes less well defined as the probability of links being active decreases. It is observed that, for all methods, the error values range between $10^{-3}$ and $10^{-4}$. Furthermore, we observe that the number of iterations required to reach the dip region is higher in UAF as compared to TSAU with BAF registering the lowest errors for all nodes. this is due to the fact that the nodes are more tightly synchronized in UAF and TSAU as compared to BAF. Also for all methods, it is observed that, the number of iterations required to reach the dip error remains nearly the same irrespective of the link availability.

\subsection{Simulation on a Network with Malicious Node}
\begin{figure}[ht!]
	\centering
	\includegraphics[width=0.50\textwidth]{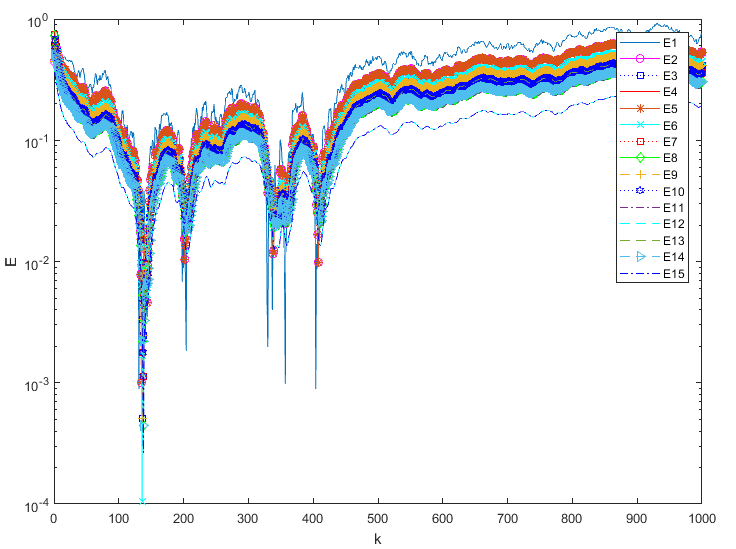}
	\caption[Node Error profile for the Test Network with Malicious Node for TSAU]{Node Error profile for the Test Network with Malicious Node for TSAU}
	\label{fig:mal_tsau}
\end{figure}

\begin{figure}[ht!]
	\centering
	\includegraphics[width=0.50\textwidth]{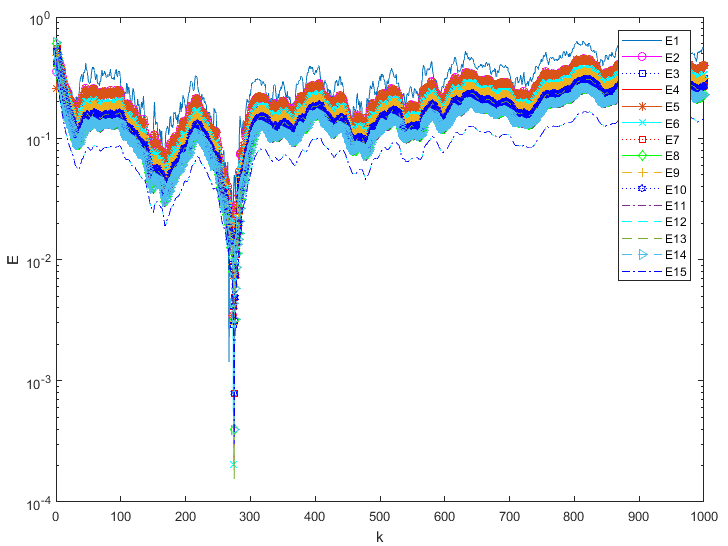}
	\caption[Node Error profile for the Test Network with Malicious Node for UAF]{Node Error profile for the Test Network with Malicious Node for UAF}
	\label{fig:mal_uaf}
\end{figure}

\begin{figure}[ht!]
	\centering
	\includegraphics[width=0.50\textwidth]{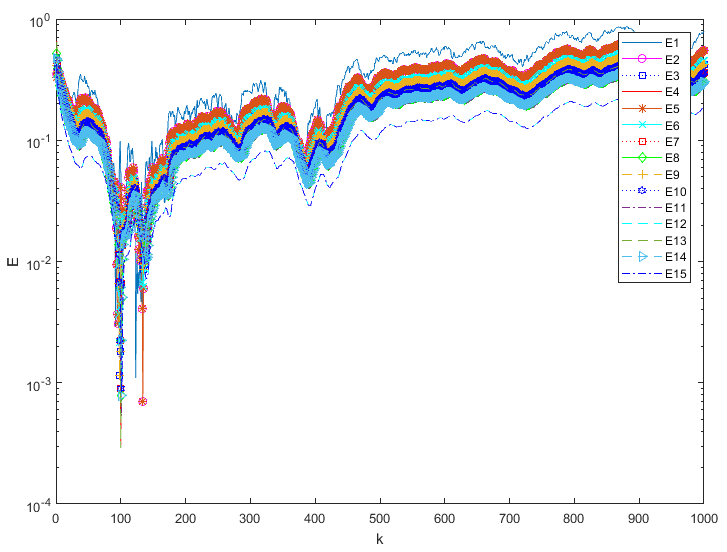}
	\caption[Node Error profile for the Test Network with Malicious Node for BAF]{Node Error profile for the Test Network with Malicious Node for BAF}
	\label{fig:mal_baf}
\end{figure}

In this section, we investigate the behavior of each of the suggested protocols when one network node, with the objective of disrupting the operation of the wireless sensor network, injects random noise instead of true time estimates into the network. This node is named, a \textit{malicious} node~\cite{pires2004malicious}. This case normally occurs when hackers or saboteurs try to compromise the operation or setup of WSNs~\cite{atassi2013malicious}. In our simulation test for each suggested protocol, the colored random noise is injected by the node furthest away from the gateway node into the network by transmitting values of this noise process at each communication instant. The statistics of this noise process is described in \cite{kasdin1995discrete}. The generated noise signal has a unity standard deviation and zero mean value and power spectral density slope of $-6 dB/octave$. Colored noise is considered here rather than white noise because colored noise is less detectable as compared to white noise and hence represents a preferred choice for saboteurs \cite{atassi2013malicious}. 
Figures \ref{fig:mal_tsau}, \ref{fig:mal_uaf} and \ref{fig:mal_baf} show the simulation results for TSAU, UAF and BAF respectively on a 16 node grid network  given in Figure \ref{fig:topps}a with a malicious node. We observe that the dip characteristic appears in the error profiles of all the methods although not well defined. For all proposed methods, we observe an average minimum error in $\mathbb{E}^{dip}_{min}$ in the $10^{-4}$ range. For TSAU and UAF, the variances $\mathbb{V}_{k^{dip}_{min}}$ for all network nodes are 3.84 and 4.29 iterations respectively as compared to that of BAF occurring at 165.57. It is also observed that, UAF has a relatively higher convergence time, $k^{dip}_{min}$ of 274 when a malicious node exist in the network as compared to TSAU and BAF with respective values of 137 and 106 iterations. Table \ref{tab:mal} summarizes these performance metrics.

\begin{table}[ht!]
	\centering
	\caption{Performance of TSAU, UAF and BAF in a Network with Malicious Node}
	\begin{tabular}{ccccccc}
		\toprule
		\textbf{Protocols} & \multicolumn{2}{c}{\textbf{TSAU}} & \multicolumn{2}{c}{\textbf{UAF}} & \multicolumn{2}{c}{\textbf{BAF}} \\
		Nodes & $k^{dip}_{i}$ & Error  & $k^{dip}_{i}$ & Error  & $k^{dip}_{i}$ & Error  \\
		\midrule
		N1   & 131  & 8.84E-04 & 267  & 1.41E-03 & 123  & 1.11E-03 \\
		N2   & 135  & 1.00E-03 & 273  & 3.42E-03 & 134  & 6.99E-04 \\
		N3   & 137  & 5.09E-04 & 274  & 2.92E-03 & 98   & 1.15E-03 \\
		N4   & 138  & 6.28E-04 & 275  & 1.55E-04 & 100  & 2.90E-04 \\
		N5   & 135  & 1.00E-03 & 273  & 3.42E-03 & 134  & 6.99E-04 \\
		N6   & 136  & 1.05E-04 & 274  & 2.05E-04 & 98   & 2.14E-03 \\
		N7   & 138  & 1.13E-03 & 275  & 7.92E-04 & 100  & 8.97E-04 \\
		N8   & 138  & 4.43E-04 & 275  & 3.93E-04 & 101  & 7.84E-04 \\
		N9   & 137  & 5.09E-04 & 274  & 2.92E-03 & 98   & 1.15E-03 \\
		N10  & 138  & 1.13E-03 & 275  & 7.92E-04 & 100  & 8.97E-04 \\
		N11  & 138  & 3.50E-04 & 275  & 5.12E-04 & 101  & 5.32E-04 \\
		N12  & 138  & 2.64E-04 & 275  & 3.01E-04 & 101  & 8.34E-04 \\
		N13  & 138  & 6.28E-04 & 275  & 1.55E-04 & 100  & 2.90E-04 \\
		N14  & 138  & 4.43E-04 & 275  & 3.93E-04 & 101  & 7.85E-04 \\
		N15  & 138  & 2.64E-04 & 275  & 3.02E-04 & 101  & 8.34E-04 \\
		\midrule
		\multicolumn{7}{c}{Statistics} \\
		\midrule
		Mean & 137  & 6.20E-04 & 274  & 1.21E-03 & 106  & 8.74E-04 \\
		Minimum & 131  & 1.05E-04 & 267  & 1.55E-04 & 98   & 2.90E-04 \\
		Maximum & 138  & 1.13E-03 & 275  & 3.43E-03 & 134  & 2.15E-03 \\
		Variance & 3.84 & 1.12E-07 & 4.29 & 1.63E-06 & 165.57 & 1.94E-07 \\
		\bottomrule
	\end{tabular}%
	\label{tab:mal}%
\end{table}%

\section{Real-Time Experimental Evaluation}
\label{Real-Time Experimental Evaluation}
To assess the performance of our synchronizer in real time settings, experimental evaluations are carried out on a testbed of MicaZ sensor motes. Further experiments are carried out on Flooding Time Synchronization Protocol(FTSP) and Flooding-based Proportional-Integral Synchronization protocol (FloodPISync) found in the literature \footnote{The source codes for FTSP and FloodPISync are publicly available in the GitHub TinyOS repoistory}. FTSP and FloodPISync are \textit{flooding-based centralized} protocols where time synchronization is done by synchronizing all network nodes to a reference node. In FTSP the estimation of the reference node clock is done using linear regression whereas FloodPISync, it is done by adjusting the node drifts and offsets using a proportional-integral controller analogy.

\subsection{Test Networks}
\begin{figure}[htbp!]
	\centering
	\subfloat[Grid Topology]{%
		\includegraphics[clip,width=0.4\columnwidth]{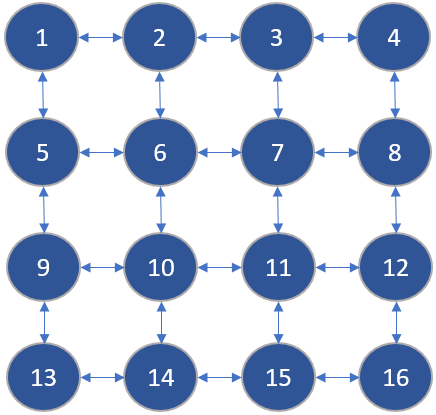}%
	}
	
	\subfloat[Line Topology]{%
		\includegraphics[clip,width=0.9\columnwidth]{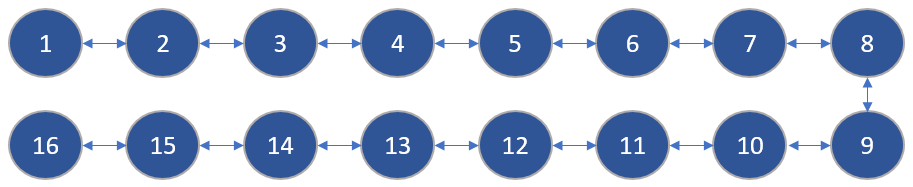}%
	}
	\caption[Node Distribution of Sensor Node for Practical Experiments]{Node Distribution of Sensor Node for Practical Experiments}
	\label{fig:topps}
\end{figure}

In our experiments, we utilize a line and grid topology of 16 nodes shown in Figure ~\ref{fig:topps}. Experiments on line topologies allow for the scalability of protocols to be observed since the performance of flooding based time synchronization protocols degrade with increase in network diameter ~\cite{lenzen2015pulsesync} whereas experiments on the grid topology give an indication of the performance of time synchronization protocols when subjected some adverse network conditions like contention, congestion and increased packet collisions ~\cite{yildirim_external_2014}. It should be noted here that, in order to realize these network configurations for our experiments, network nodes are forced to accept packets from only some nodes.

\subsection{Hardware Platform}
Practical experiments are done on a network platform comprising of MicaZ nodes manufactured by Memsic Inc. These nodes are instrumented with a 7.37MHz 8-bit Atmel Atmega128L microcontroller. They are also equipped with 4kB RAM, 128kB program flash and Chipcon CC2420 radio chip which provides a 250 kbps data rate at 2.4 GHz frequency. The 7.37MHz quartz oscillator on the MicaZ board is employed as the clock source for the timer used for timing measurements. This timer operates at 1/8 of that frequency and thus each timer tick occurs at approximately every 921 kHz (approx. 1 $\mu s$). TinyOS-2.1.2 installed on Ubuntu Linux Distribution 14 is used a the base operating system for all experimental work.

The CC2420 transceiver on the MicaZ board employed has the capability to timestamp synchronization packets at MAC layer with the timer used for timing measurements. This is a well-known method that increases the quality of time synchronization by reducing the effect of non-deterministic error sources arising from communication \cite{lee_low-complexity_2016}. Packet level time synchronization interfaces provided by TinyOS are utilized to timestamp synchronization messages at the MAC layer ~\cite{Maroti_ttp}.

\subsection{Testbed Setup}
The experiments are carried out by configuring a number of nodes in a line and grid topologies with one of the nodes configured as the gateway node and the rest of the nodes acting as slaves, exchanging their time synchronization packets and running the time synchronization algorithm. One node is programed as the reference node which periodically transmit query packets to the other nodes to trigger the synchronization protocol. Another node is programmed as the sink or base-station and collects all time-sync packets onto a computer for review.

\subsection{Experiment Parameters}
In our experiments, a beacon period, $B$ of 30 seconds was used for all protocols. The least-squares regression table used for FTSP is kept at a capacity of 8 elements as used for the comparative work in \cite{yildirim2014time}. The experimental parameters for FloodPISync, $\beta$, $\alpha_{max}$ and $e_{max}$ are kept respectively at 1, $3.33 \times 10^{-8}$ and 6000 $ticks$ as used for comparisons in ~\cite{yildirim2018adaptive}. Since it takes a number of pooling cycles for FTSP to elect the root node, the protocol is modified to have a fixed root node. Node '1' is used as the reference node for the FTSP and FloodPISync experiments and used as the gateway node for TSAU, UAF and BAF. A TMicro timer ticking periodically at an interval of 1000\footnote{i.e., $\Delta$ of 1 $ms$} is used for the gateway node in the experiments for TSAU, UAF and BAF.  When each experiment commences, network nodes are switched on randomly within 2 minutes of operation with each experiment taking up to about 3.8 hours.

\subsection{Energy Consumption and Memory Requirements}
In this section we present an evaluation of the performances of FTSP, FloodPISync, TSAU, UAF and BAF in terms energy consumption and memory requirements. For a time synchronization protocol to operate efficiently, it must conserve resources which are deemed scarce for WSN nodes. The Random Access Memory (RAM) is the memory used to store node information obtained from the network during the execution of the protocol and therefore determines the main memory requirements of the protocol \footnote{The Memsic MicaZ nodes are equipped with 4kB RAM}. This resource must be conserved especially in distributed protocols where the retention of neighborhood clock information is necessary for effective operation. For each node to store the main protocol application source code and the auxiliary codes and interfaces required to carryout synchronization, the Read-Only Memory is needed, which is also a limited resource and must be conserved. The core heuristics and main applications of the protocol must be effective to synchronize all network nodes but simple enough to fit on the ROM\footnote{Each Memsic MicaZ node is equipped with 128kB program flash memory}. Since most wireless sensor nodes perform other tasks, like routing, clustering, congestion control and traffic management, the time synchronization protocol should occupy only a small fraction of the available ROM space. Apart from conserving the RAM and ROM, each protocol must also conserve the energy consumption of network nodes. Crucial for energy consumption is the number of communication cycles, the CPU overhead \footnote{The CPU overhead is the time in \textit{ticks} taken for a recently received synchronization message to be processed and used to update the clock of a node. It should be noted that $1 \ tick = 1 \ \mu s$.} and the length of synchronization messages. For the number of communication cycles, the higher the frequency at which nodes communicate, the higher the amount of information needed to be exchanged to achieve synchronization and the more energy is expended in the transmission and reception of synchronization messages. Also the higher the CPU overhead, the higher the energy needed to process and update the time synchronization information. The length of the message is also a crucial determinant in the time taken to transmit, receive and process synchronization messages. In Table ~\ref{tab:memory}, we present the CPU overhead, the message length and the main memory (RAM) and flash momory (ROM) requirements for FTSP, FloodPISync, TSAU, UAF and BAF.

\begin{figure}[ht]
	\centering
	\subfloat[FTSP and FloodPISync Payload Fields]{%
		\includegraphics[clip,width=\columnwidth]{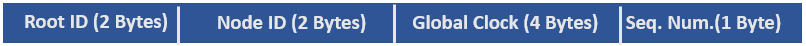}%
	}
	
	\subfloat[BAF Payload Fields]{%
		\includegraphics[clip,width=1.0\columnwidth]{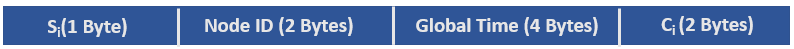}%
	}
	
	\subfloat[UAF Payload Fields]{%
		\includegraphics[clip,width=0.75\columnwidth]{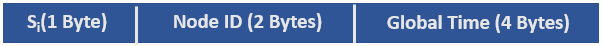}%
	}
	
	\subfloat[TSAU Payload Fields]{%
		\includegraphics[clip,width=0.6\columnwidth]{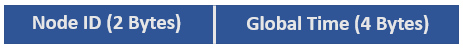}%
	}
	
	\caption[TimeSync Packet Payload Field Descriptions for FTSP, FloodPISync, TSAU, UAF and BAF]{TimeSync Packet Payload Field Descriptions for FTSP, FloodPISync, TSAU, UAF and BAF}
	\label{fig:pkftsp}
\end{figure}
In the operation of FTSP, when a new synchronization packet is received, FTSP is designed to store the current clock information in an 8 element regression table and conduct least square regression computations involving several float point divisions and multiplications ~\cite{maroti_flooding_2004}, hence is observed to have the high CPU overhead of 5440 $ticks$. FloodPISync is lightweight and carry out few additions and multiplications with almost no float point divisions and are therefore observed to each consume a significantly reduced computation time or CPU overhead of 145 $ticks$. TSAU carries out an averaging calculation in each cycles and also maintains the parameter, $UpdateTime$ which regulates the sequential activation. TSAU has a CPU overhead of 133. For UAF, the mathematical computations involves just the additions and floating point division used in calculating the average and hence has the lowest CPU overhead of 133 $ticks$, whereas in BAF the CPU overhead increases by 29 $ticks$ due to the added logical operations involving the counter variable, $c_i$ \footnote{$c_i$ is the variable included to ensure automatic regulation of the asynchronous bidirectional cycling}.\\

\begin{table}[htbp]
	\centering
	\caption{Memory Requirements, CPU Overhead, and Synchronization Message Length of FTSP, FloodPISync, TSAU, UAF and BAF}
	\hspace*{-1.8em}
	\begin{tabular}{lrrrrr}
		\toprule
		\textbf{Protocols} & \multicolumn{1}{c}{FTSP} & \multicolumn{1}{c}{FloodPISync} & \multicolumn{1}{c}{TSAU} & \multicolumn{1}{c}{UAF} & \multicolumn{1}{c}{BAF} \\
		\cmidrule{2-6}    \textbf{CPU Overhead (ticks)} & 5440 & 145  & 141  & 133  & 162 \\
		\textbf{Message Length (bytes)} & 9    & 9    & 4    & 7    & 9 \\
		\textbf{Main Memory (bytes)} & 52   & 16   & 16   & 14   & 12 \\
		\textbf{Flash Memory (bytes)} & 17974 & 17974 & 13510 & 330  & 12282 \\
		\bottomrule
	\end{tabular}%
	\label{tab:memory}%
\end{table}%

As shown in Figure ~\ref{fig:pkftsp} (a), for FTSP and FloodPISync, each synchronization payload is composed of 2 $bytes$ reference or root node ID, 2 $bytes$ node ID of the particular node sending the packet, 4 $bytes$ current clock information and 1 $byte$ sequence number, making a total message length of 9 $bytes$ as shown in Table ~\ref{tab:memory}. TSAU payload has a length of 6 $bytes$ composed of node ID of the particular node sending the packet and 4 $bytes$ global time as shown in Figure ~\ref{fig:pkftsp} (d). For UAF, each synchronization payload shown in Figure ~\ref{fig:pkftsp} (c) is composed of 2 $bytes$ node ID of the particular node sending the packet, 4 $bytes$ current clock information and 1 $bytes$ for the binary-status bit variable, $s_i$, making a total message length of 7 $bytes$. The added counter variable, $c_i$ is the only difference between BAF and UAF packets and hence BAF has a total message length of 9 $bytes$ similar to FTSP and FloodPISync. Based on the clock model presented by Equation 2.2 in ~\cite{yildirim2018adaptive} which is employed in the source code of all these protocols, each protocol is required to store a hardware clock of 4 $bytes$, a logical clock of 4 $bytes$ and a rate multiplier of 4 $bytes$. In addition to this, FTSP allocates 40 $bytes$ for its regression table ~\cite{maroti_flooding_2004} whereas FloodPISync allocates 4 $bytes$ for storing an error parameter ~\cite{yildirim2018adaptive}.  For TSAU, UAF and BAF, no memory allocation is required for the storage of neighboring node(s) or gateway node clocks, but 2 $bytes$ is allocated for the counter variable, $c_i$ in BAF. Therefore as shown in Table ~\ref{tab:memory}, the main memory (RAM) required for the execution and operations of FTSP, FloodPISync, TSAU, UAF and BAF are 52 $bytes$, 16 $bytes$,12 $bytes$, 12 $bytes$ and 14 $bytes$ respectively. Hence, TSAU, UAF and BAF outperform all the other reported protocols in terms of memory requirements. For the flash memory (ROM), FTSP requires 17947 $bytes$, FloodPISync requires 16352 $bytes$, TSAU requires 13510 $bytes$, UAF requires 13320 $bytes$ and BAF requires 12282 $bytes$.

\begin{table}[ht]
	\centering
	\caption{Current Consumption by MicaZ Nodes \cite{polastre2005telos}}
	\begin{tabular}{lc}
		\toprule
		\textbf{Operation} & \textbf{MicaZ} \\
		\midrule
		Minimum Voltage & 2.7 V \\
		Standby & 27.0 $\mu A$ \\
		MCU Idle & 3.2 $mA$ \\
		MCU Active & 8.0 $mA$\ \\
		MCU + Radio RX & 23.3 $mA$ \\
		MCU + Radio TX (0 dBm) & 21.0 $mA$ \\
		\bottomrule
	\end{tabular}%
	\label{tab:power}%
\end{table}%

In order to calculate the total energy consumed, we need to combine the total required energy for transmission, reception and processing. For the Memsic MicaZ nodes employed in this work, the minimum voltage, $v_{min}$ required, the current consumed in transmission, $i_{TX}$, the current consumed in reception, $i_{RX}$ and the current consumed by the microcontroller in active mode, $i_{MCU}$ are given in Table ~\ref{tab:power}. Given a default 11 $bytes$ header and 7 $bytes$ footer for TinyOS packets, the total packet length, $L$ of FTSP, FloodPISync and BAF is 27 $bytes$ each, that of TSAU is 24 $bytes$ each, and that of UAF, 25 $bytes$. Given a CPU overhead, $c$ and a data rate, $R$ of 250 kbps  \footnote{Chipcon CC2420 radio chip  provides a 250 kbps data rate at 2.4 GHz frequency.}, the total energy, $E_{T}$ require for each protocol can given as;
\begin{equation}\label{energy}
E_{T} = c\times i_{MCU}\times v_{min} + \frac{L}{R}\times i_{TX}\times v_{min} + \frac{L}{R}\times i_{RX}\times v_{min}
\end{equation}

Based on (\ref{energy}), the energy consumptions required for FTSP, FloodPISync, TSAU, UAF and BAF are given respectively as 130.4 $\mu J$, 16.1 $\mu J$, 14.53 $\mu J$, 14.8 $\mu J$ and 16.4 $\mu J$. Hence we observe that TSAU consumes least power but has nearly a similar consumption with UAF followed by the PISync protocols then BAF. EGTSP consumes the highest power among all the protocols. We see here that our proposed protocols consume about 9 times less than FTSP but approximately similar consumption as the PISync protocol.

\subsection{Local and Global Synchronization Errors}
To evaluate the synchronization accuracy for each protocol, the differences in clock values of network nodes has to be observed. We accomplish this by collecting all nodes' clock values at each communication instant, $k$. Network-wide synchronization error at each communication instant is then calculated using the \textit{maximum global synchronization error}, $E^{G}_{max}$ and the \textit{average global synchronization error}, $E^{G}_{avg}$ which presents a measure of the global synchronization accuracy for all nodes given by (\ref{mG}) and (\ref{aG}) respectively ~\cite{yildirim2018adaptive}. Further, we observe the synchronization error of nodes with respect to their neighbors by using the \textit{maximum local synchronization error}, $E^{L}_{max}$ and the \textit{average local synchronization error}, $E^{L}_{avg}$ which given  respectively by (\ref{mL}) and (\ref{aL}) ~\cite{yildirim2014time}. An important note here is that, since our suggested synchronization ceases update at the dip region, it is fair to compare the above described parameters with the global and local errors in the dip region for our developed method.  The time taken for the network to synchronize to a global time, i.e. convergence time is also compare. For TSAU, UAF and BAF, the convergence time is the same as the number of communication cycles needed to reach the dip region represented by the parameter, $k^{dip}_{min}$.

\begin{equation}\label{mG}
E_{max}^{G} = \max\limits_{i,j\in V} (t_i(k) - t_j(k))
\end{equation}
\begin{equation}\label{mL}
E_{max}^{L} = \max\limits_{i\in V,j\in E} (t_i(k) - t_j(k))
\end{equation}
\begin{equation}\label{aG}
E_{avg}^{G} = \frac{1}{N} \sum^{}_{i \in V}  \max\limits_{i,j\in V} (t_i(k) - t_j(k))
\end{equation}
\begin{equation}\label{aL}
E_{avg}^{L} = \frac{1}{N} \sum^{}_{i \in V} \max\limits_{i\in V,j\in E} (t_i(k) - t_j(k))
\end{equation}

Where $V$ and $E$ are the global network vertex set and the neighborhood graph vertex set respectively and $N$ is the number of nodes in the network.

\subsubsection{Comparison: Synchronization Error for Grid Topology}
\begin{figure}[ht]
	\centering
	\includegraphics[width=0.5\textwidth]{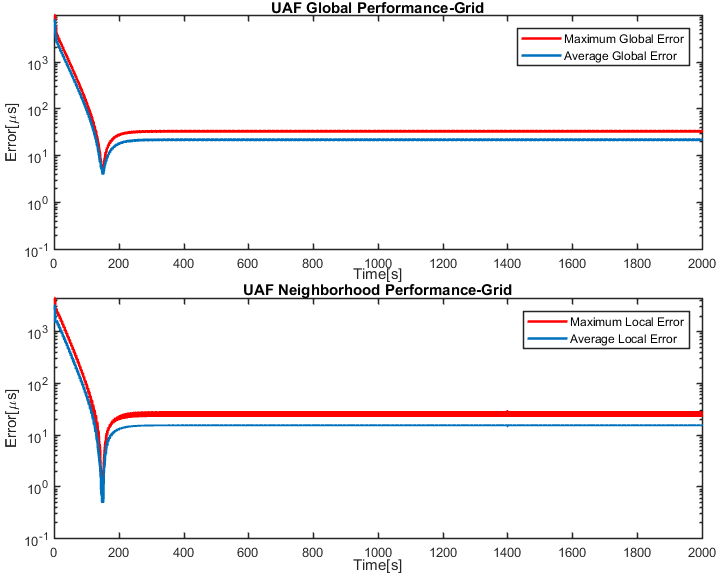}
	\caption[UAF Neighborhood and Global Synchronization Error for Grid Topology]{UAF Neighborhood and Global Synchronization Error for Grid Topology}
	\label{fig:taufGrid}
\end{figure}

\begin{figure}[ht]
	\centering
	\includegraphics[width=0.5\textwidth]{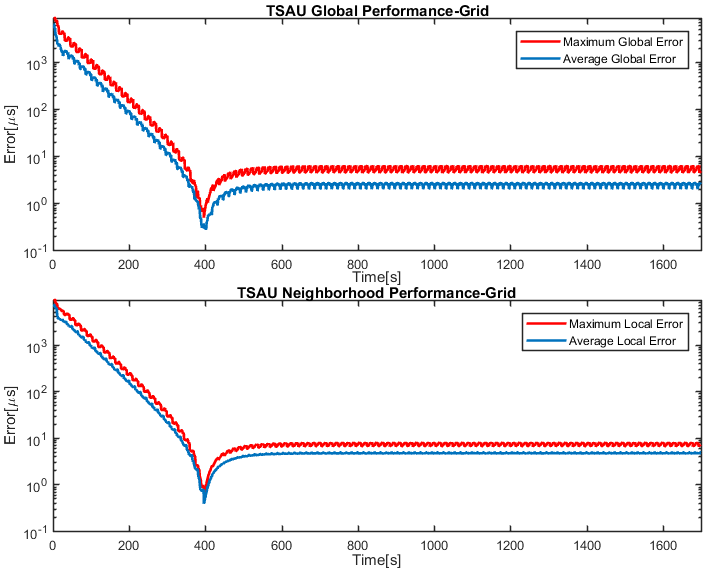}
	\caption[TSAU Neighborhood and Global Synchronization Error for Grid Topology]{TSAU Neighborhood and Global Synchronization Error for Grid Topology}
	\label{fig:tsapGrid}
\end{figure}

\begin{figure}[ht]
	\centering
	\includegraphics[width=0.5\textwidth]{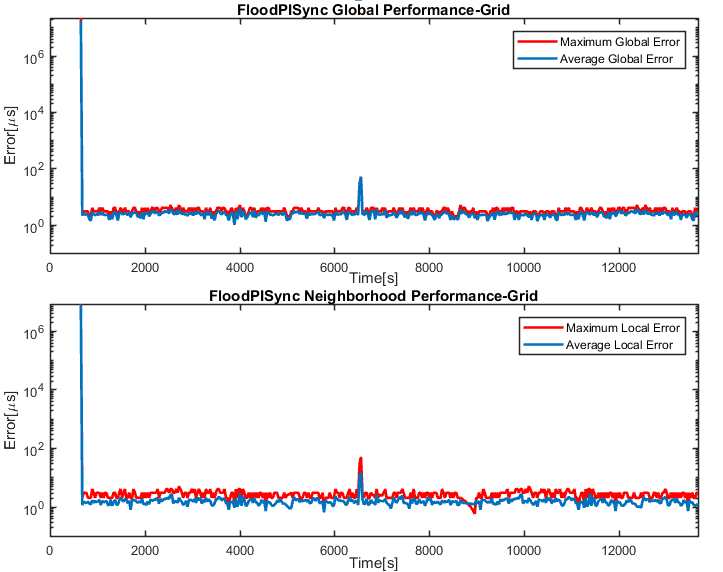}
	\caption[FloodPISync Neighborhood and Global Synchronization Error for Grid Topology]{FloodPISync Neighborhood and Global Synchronization Error for Grid Topology}
	\label{fig:floodpisyncGrid}
\end{figure}

\begin{figure}[ht]
	\centering
	\includegraphics[width=0.5\textwidth]{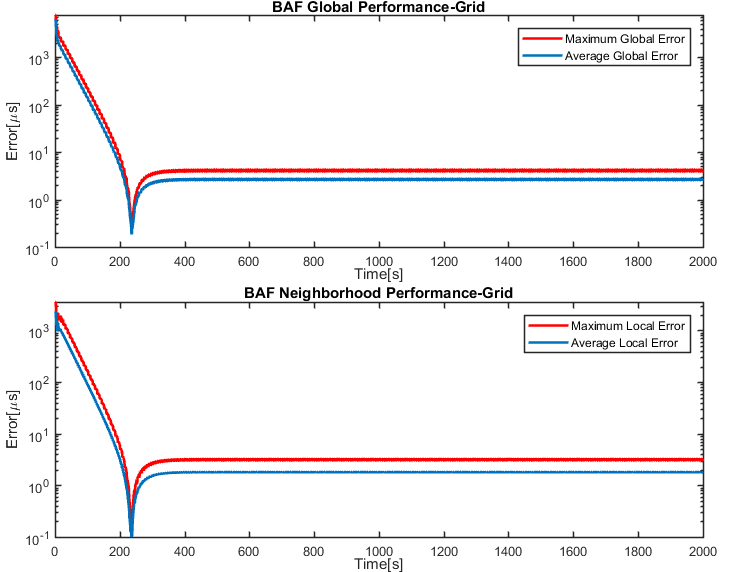}
	\caption[BAF Neighborhood and Global Synchronization Error for Grid Topology]{BAF Neighborhood and Global Synchronization Error for Grid Topology}
	\label{fig:bafpGrid}
\end{figure}

\begin{figure}[ht]
	\centering
	\includegraphics[width=0.5\textwidth]{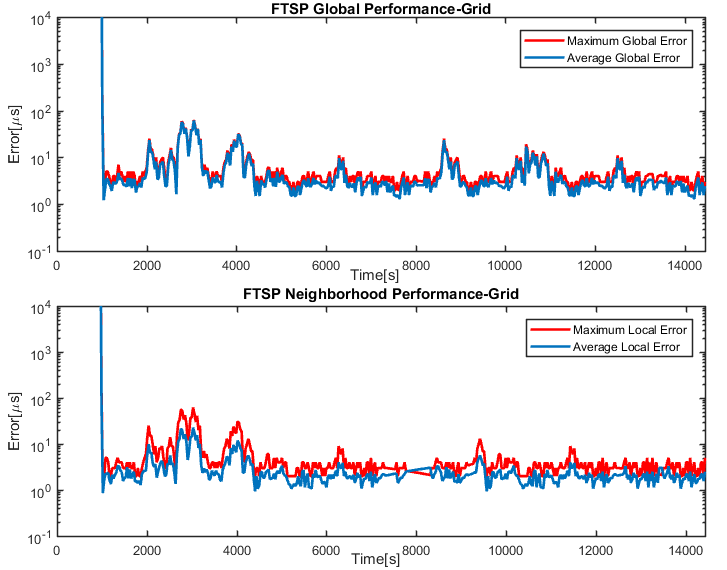}
	\caption[FTSP Neighborhood and Global Synchronization Error for Grid Topology]{FTSP Neighborhood and Global Synchronization Error for Grid Topology}
	\label{fig:ftspGrid}
\end{figure}

In this section we compare the performance of each protocol in terms of local and global errors for the grid topology. The maximum and average, local and global errors are presented for FTSP, FloodPISync, TSAU, UAF and BAF by Figures ~\ref{fig:ftspGrid}, ~\ref{fig:floodpisyncGrid}, ~\ref{fig:tsapGrid}, ~\ref{fig:taufGrid} and ~\ref{fig:bafpGrid} respectively. The critical values for these metrics including the convergence time are summarized in Table ~\ref{tab:compGrid}. First we observe that, the global and local error curves for TSAU, UAF and BAF show a more smooth characteristic as compared to FTSP and FloodPISync.

This might stem from the fact that in our methods of time update, clock estimation is not done by an independent estimation of clock skew rate and offset like in these other protocols but is done by directly using the global logical time values. Although individual estimation of skew rate and offset is widely reported in a myriad of synchronization protocols, for our method of synchronization, the skew rate and offset would still converge to a consensus value as long as for any network node, $i$, there exist a spanning path from this node to the gateway node.

Secondly, we observe that the \textit{dip} characteristics of the algorithm appears in all global and local errors. With a well defined minimum error our proposed method is expected to locate the transient and stop update. This is a significant advantage for the proposed synchronization protocol because unlike the other protocols where nodes are continuously, communicating and estimating the global time through drift and offset compensation, nodes running on TSAU, UAF and BAF only need to identify the transient dip, stop updates and sleep to conserve a lot of energy.

From the results we observe that, the global and local errors for BAF far out perform all the other protocols as shown in Table \ref{tab:compGrid}. This is followed by TSAU then UAF. We further observe from Table \ref{tab:compGrid} that, FTSP outperforms FloodPISync on the grid topology in terms all the errors registering an average local error on 4.1$\mu s$ as compared to 6.2$\mu s$ for FloodPISync.

\begin{table}[ht]
	\centering
	\caption{Global and Local Error Comparison-Grid Topology}
	\begin{tabular}{lrrrrr}
		\toprule
		\multicolumn{1}{c}{\textbf{Protocols}} & \multicolumn{1}{c}{FTSP} & \multicolumn{1}{c}{FloodPISync} & \multicolumn{1}{c}{TSAU} & \multicolumn{1}{c}{UAF} & \multicolumn{1}{c}{BAF} \\
		\cmidrule{1-1}    \textbf{Max. Global($\mu s$)} & \multicolumn{1}{c}{8.0} & \multicolumn{1}{c}{13.0} & 0.73 & 4.00 & 0.42 \\
		\textbf{Avg. Global($\mu s$)} & \multicolumn{1}{c}{6.9} & \multicolumn{1}{c}{11.6} & 0.47 & 3.80 & 0.20 \\
		\textbf{Max. Local($\mu s$)} & \multicolumn{1}{c}{8.0} & \multicolumn{1}{c}{13.0} & 0.53 & 1.50 & 0.23 \\
		\textbf{Avg. Local($\mu s$)} & \multicolumn{1}{c}{4.1} & \multicolumn{1}{c}{6.2}  & 0.35 & 0.47 & 0.14 \\
		\textbf{Conv. Time($s$)} & \multicolumn{1}{c} {$\simeq$900}  & \multicolumn{1}{c} {$\simeq$750}  & \multicolumn{1}{c}{$\simeq$400}  & \multicolumn{1}{c}{$\simeq$155}  & \multicolumn{1}{c}{$\simeq$230} \\
		\bottomrule
	\end{tabular}%
	\label{tab:compGrid}%
\end{table}%

On the convergence time, TSAU, UAF and BAF also out performed all the other protocols.  Our proposed methods are all decentralized and communication is done only with 1-hop neighbors therefore a lower convergence time is expected. Since for TSAU only one node updates at a time, we observe that it takes about twice more time for the network to converge as compared to UAF and BAF as shown in Table \ref{tab:compGrid}.

\subsubsection{Comparison: Synchronization Error for Line Topology}
\begin{figure}[ht]
	\centering
	\includegraphics[width=0.5\textwidth]{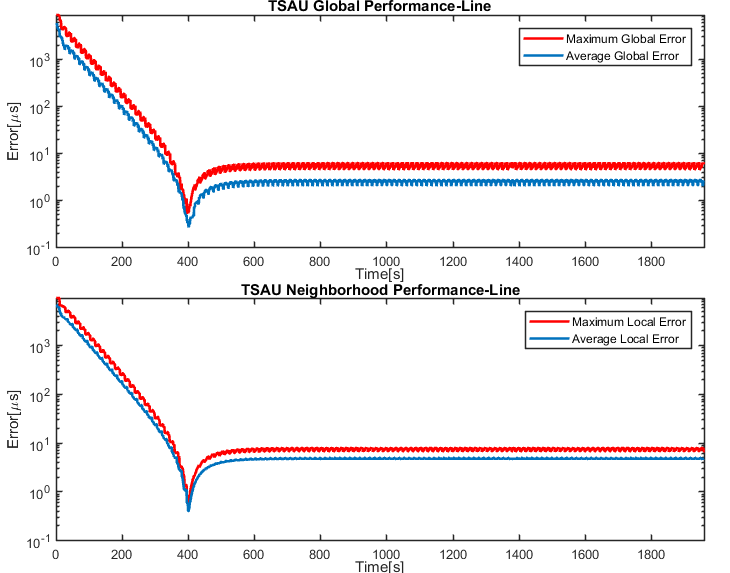}
	\caption[TSAU Neighborhood and Global Synchronization Error for Line Topology]{TSAU Neighborhood and Global Synchronization Error for Line Topology}
	\label{fig:tsapLine}
\end{figure}

\begin{figure}[ht]
	\centering
	\includegraphics[width=0.5\textwidth]{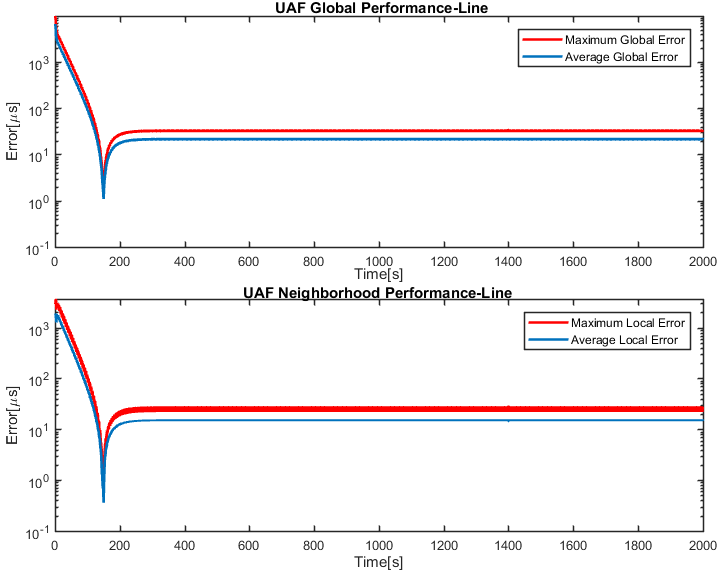}
	\caption[UAF Neighborhood and Global Synchronization Error for Line Topology]{UAF Neighborhood and Global Synchronization Error for Line Topology}
	\label{fig:taufLine}
\end{figure}

\begin{figure}[ht]
	\centering
	\includegraphics[width=0.5\textwidth]{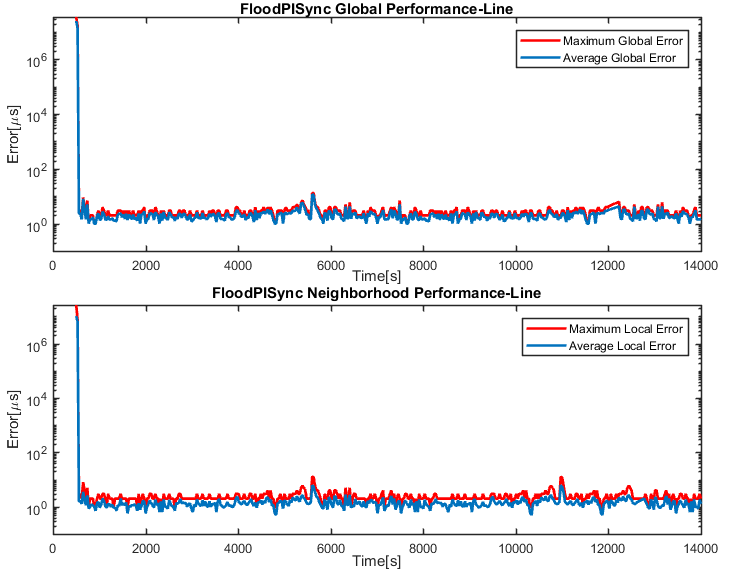}
	\caption[FloodPISync Neighborhood and Global Synchronization Error for Line Topology]{FloodPISync Neighborhood and Global Synchronization Error for Line Topology}
	\label{fig:floodpisyncLine}
\end{figure}

\begin{figure}[ht]
	\centering
	\includegraphics[width=0.5\textwidth]{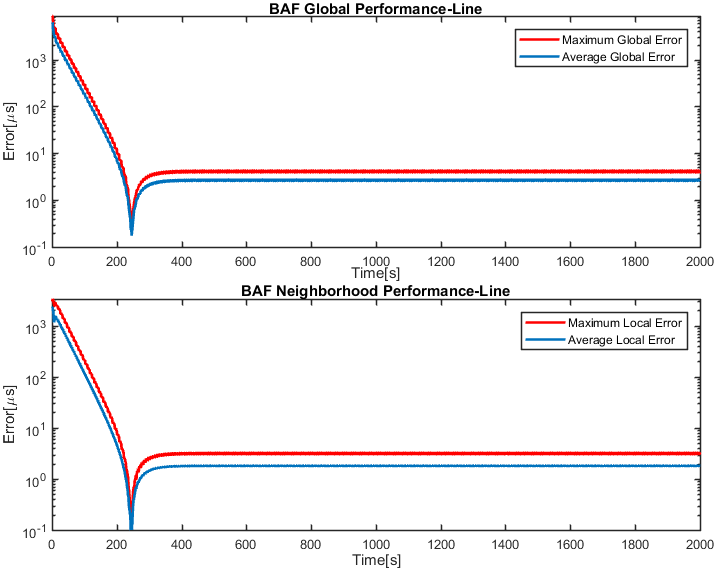}
	\caption[BAF Neighborhood and Global Synchronization Error for Line Topology]{BAF Neighborhood and Global Synchronization Error for Line Topology}
	\label{fig:bafpLine}
\end{figure}

\begin{figure}[ht]
	\centering
	\includegraphics[width=0.5\textwidth]{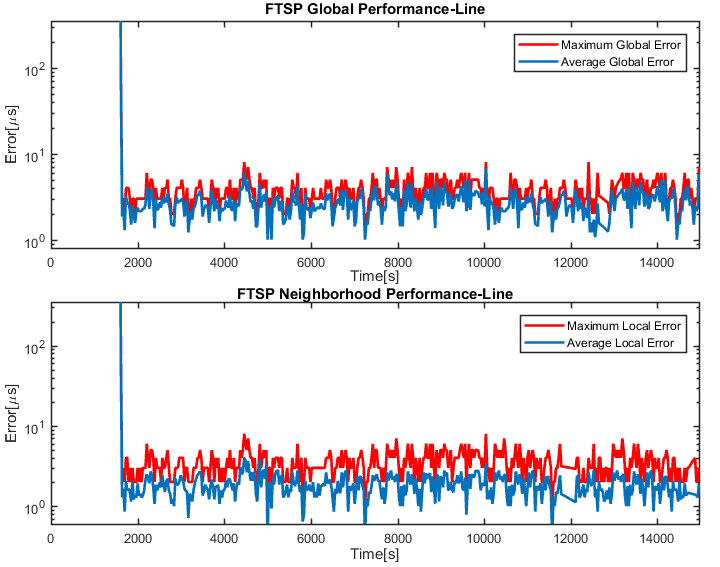}
	\caption[FTSP Neighborhood and Global Synchronization Error for Line Topology]{FTSP Neighborhood and Global Synchronization Error for Line Topology}
	\label{fig:ftspLine}
\end{figure}

The results for global and local performances of the protocols in a line topology are presented in this section. The maximum and average, local and global errors are presented for FTSP, FloodPISync, TSAU, UAF and BAF by Figures ~\ref{fig:ftspLine}, ~\ref{fig:floodpisyncLine}, ~\ref{fig:tsapLine}, ~\ref{fig:taufLine} and ~\ref{fig:bafpLine} respectively and Table ~\ref{tab:compLine} summarizes these results including the convergence time.

For TSAU, UAF and BAF, the characteristics of the error curves are observed to be similar to those of the grid topology but almost all the global and local errors have surprisingly lower values as compared to those in the grid  topology. For TSAU we observed all global and local error values to be below 0.7 $\mu s$ as compared the grid topology where values go as high as 4$\mu s$. This is understandable because TSAU operate one node at a time so a structured line networks fit its operation as compared to the grid topology. But TSAU still recorded higher error values compared to UAF and BAF. BAF still performed best in terms of local and global error followed by UAF.

For FTSP, we observe a graph with high overshoots and undershoots as compared to the grid error curves.Also we note that, BAF has the lowest local and global skews as compared to all other schemes.  All the other protocols also registers in the tens compared to values less than one in TSAU, UAF and BAF. Hence, once the stopping criterion is able to stop node updates in the dip region, our proposed protocol will outperformed all these protocols by a huge margin and therefore at worst, on average, after convergence is achieve in the dip region, it will take about 200 times the same time for our proposed methods to drift to the same level as FTSP and FloodPISync.

\begin{table}[h!]
	\centering
	\caption{Global and Local Error Comparison-Line Topology}
	\begin{tabular}{lrrrrr}
		\toprule
		\multicolumn{1}{c}{\textbf{Protocols}} & \multicolumn{1}{c}{FTSP} & \multicolumn{1}{c}{FloodPISync} & \multicolumn{1}{c}{TSAU} & \multicolumn{1}{c}{UAF} & \multicolumn{1}{c}{BAF} \\
		\cmidrule{1-1}    \textbf{Max. Global($\mu s$)} & \multicolumn{1}{c}{63.0} & \multicolumn{1}{c}{51.0} & 0.66 & 1.01 & 0.34 \\
		\textbf{Avg. Global($\mu s$)} & \multicolumn{1}{c}{61.5} & \multicolumn{1}{c}{50.0} & 0.30 & 2.80 & 0.18 \\
		\textbf{Max. Local($\mu s$)} & \multicolumn{1}{c}{62.0} & \multicolumn{1}{c}{50.0} & 0.57 & 0.77 & 0.21 \\
		\textbf{Avg. Local($\mu s$)} & \multicolumn{1}{c}{22.7} & \multicolumn{1}{c}{15.4} & 0.38 & 0.36 & 0.10 \\
		\textbf{Conv. Time($s$)} & \multicolumn{1}{c}{$\simeq$2150} &\multicolumn{1}{c}{$\simeq$550} & \multicolumn{1}{c}{$\simeq$400}  & \multicolumn{1}{c}{$\simeq$150}  & \multicolumn{1}{c}{$\simeq$250} \\
		\bottomrule
	\end{tabular}%
	\label{tab:compLine}%
\end{table}%

In the line topology, we see an even larger convergence time for all protocols except FloodPISync. Also TSAU, UAF and BAF have nearly similar convergence times as in the grid topology but we observe that BAF has an increased convergence time from 230$\mu s$ in the grid topology to 250 $\mu s$ in the line network. It is also observed here that, FTSP has about more than 5 time convergence time as compared to our developed methods. FloodPISync still registered a lower convergence time compared to FTSP. Hence we can infer that amongst our protocols UAF performs best in terms of convergence time.

\section{Conclusion}
\label{Conclusion}
This paper suggests a novel and practical time synchronization protocol that suits infrastructure impoverished networks such as WSNs operating in harsh environments. The synchronizer is constructed using concepts from switched systems, consensus control and pattern recognition. The approach clearly demonstrates that resilient, accurate and energy efficient synchronization
of decentralized networks that is based on single hop asynchronous communication is both possible and practical. It is shown that the suggested synchronization protocol can be realized in different ways where each realization has its own distinctive features. The three realizations
suggested in this paper are called: Timed Sequential Asynchronous Update (TSAU), Unidirectional Asynchronous Flooding (UAF) and Bidirectional Asynchronous Flooding (BAF). The protocols were intensively tested by both simulation and experiments on the MicaZ WSN platform. They are found to provide good performance even under adverse conditions such as when a node of the network is malicious or when the communication among the nodes are
not reliable. The protocols are also extensively compared to leading protocols for synchronizing WSNs and found to have favorable performance in terms of rate of convergence, energy consumption and memory usage among other things. Future work will focus on theoretical analysis and verification of the promising properties of the suggested protocols.

\ifCLASSOPTIONcaptionsoff
  \newpage
\fi

\bibliographystyle{ieeetr}
\bibliography{IEEEabrv,ref}

\end{document}